\documentclass[twocolumn,twocolappendix]{aastex701}

\usepackage{amsmath}

\shorttitle{Detection of fine-structured radio bursts during a superflare on EQ Peg}
\shortauthors{Zhanhao Zhao et al.}
\graphicspath{{./}{figures/}}

\begin{document}

\title{Detection of fine-structured radio bursts during a superflare on EQ Peg}

\author[orcid=0009-0003-8956-547X,gname=Zhanhao,sname=Zhao]{Zhanhao Zhao}
\affiliation{School of Astronomy and Space Science, Nanjing University, Nanjing 210023, People's Republic of China}
\affiliation{Key Laboratory of Modern Astronomy and Astrophysics (Nanjing University), Ministry of Education, Nanjing 210093, People's Republic of China}
\email{}

\author[orcid=0000-0003-2837-7136,gname=Xin,sname=Cheng]{Xin Cheng} 
\affiliation{School of Astronomy and Space Science, Nanjing University, Nanjing 210023, People's Republic of China}
\affiliation{Key Laboratory of Modern Astronomy and Astrophysics (Nanjing University), Ministry of Education, Nanjing 210093, People's Republic of China}
\email[show]{xincheng@nju.edu.cn}
\correspondingauthor{Xin Cheng}

\author[orcid=0000-0002-4198-2333,gname=Yuankun,sname=Kou]{Yuankun Kou}
\affiliation{School of Astronomy and Space Science, Nanjing University, Nanjing 210023, People's Republic of China}
\affiliation{Key Laboratory of Modern Astronomy and Astrophysics (Nanjing University), Ministry of Education, Nanjing 210093, People's Republic of China}
\email{}

\author[orcid=0000-0001-9610-0433,gname=Guoyin,sname=Chen]{Guoyin Chen}
\affiliation{School of Astronomy and Space Science, Nanjing University, Nanjing 210023, People's Republic of China}
\affiliation{Key Laboratory of Modern Astronomy and Astrophysics (Nanjing University), Ministry of Education, Nanjing 210093, People's Republic of China}
\email{}

\author[orcid=0000-0001-8132-5357,gname=Hao,sname=Ning]{Hao Ning}
\affiliation{Institute of Frontier and Interdisciplinary Science, Shandong University, Qingdao 266237, People's Republic of China}
\affiliation{Institute of Space Sciences, Shandong University, Weihai 264209, People's Republic of China}
\email{}

\author[orcid=0000-0001-6449-8838,gname=Yao,sname=Chen]{Yao Chen}
\affiliation{Institute of Frontier and Interdisciplinary Science, Shandong University, Qingdao 266237, People's Republic of China}
\affiliation{Institute of Space Sciences, Shandong University, Weihai 264209, People's Republic of China}
\email{}

\author[orcid=0000-0003-2047-9664,gname=Baolin,sname=Tan]{Baolin Tan}
\affiliation{CAS Key Laboratory of Solar Activity, National Astronomical Observatories, Chinese Academy of Sciences, Beijing 100012, People's Republic of China}
\affiliation{Key Laboratory of Solar Activity and Space Weather, National Space Science Center, Chinese Academy of Sciences, Beijing 100190, People's Republic of China}
\affiliation{School of Astronomy and Space Science, University of Chinese Academy of Sciences, Beijing 100049, People's Republic of China}
\email{}

\author[orcid=0000-0002-5093-5088,gname=Keping,sname=Qiu]{Keping Qiu}
\affiliation{School of Astronomy and Space Science, Nanjing University, Nanjing 210023, People's Republic of China}
\affiliation{Key Laboratory of Modern Astronomy and Astrophysics (Nanjing University), Ministry of Education, Nanjing 210093, People's Republic of China}
\email{}

\author[orcid=0000-0002-4978-4972,gname=Mingde,sname=Ding]{Mingde Ding}
\affiliation{School of Astronomy and Space Science, Nanjing University, Nanjing 210023, People's Republic of China}
\affiliation{Key Laboratory of Modern Astronomy and Astrophysics (Nanjing University), Ministry of Education, Nanjing 210093, People's Republic of China}
\email{}

\begin{abstract}

Superflares are frequently observed on magnetically active stars, particularly on active M-dwarfs, thereby influencing the habitability of nearby exoplanets. 
However, the causes of superflares remain unclear. 
Here, combining FAST and TESS, we detected a finely structured decimetric radio burst appearing as groups of spikes during a white-light superflare (total bolometric energy $\sim 10^{33} \ \mathrm{erg}$) on the M-dwarf binary EQ Peg on September 8th, 2022. 
The most significant ones appeared near the peak time of the optical flare and lasted for about $100 \ \mathrm{s}$. 
The radio spikes present extremely high brightness temperatures ($\gtrsim 10^{13} \ \mathrm{K}$) and high degrees of circular polarization ($\sim 0.84 \pm 0.16$), indicating the origin of a coherent emission process, most likely the electron cyclotron maser. 
Based on Zeeman-Doppler Imaging measurements of EQ Peg, we further locate the sources of the radio spikes, including their heights and latitudes, under the assumption of a large-scale dipolar magnetic field configuration. 
The results provide strong evidence for finely structured decimetric radio bursts generated by stellar flares and shed light on the origin of superflares. 

\end{abstract}


\keywords{
\uat{M dwarf stars}{982} --- 
\uat{Stellar physics}{1621} --- 
\uat{Stellar flares}{1603} --- 
\uat{Stellar magnetic fields}{1610} --- 
\uat{Time domain astronomy}{2109} --- 
\uat{Radio astronomy}{1338}
}

\section{Introduction}\label{section_introduction}

\indent Late-type stars, especially M-dwarfs, are thought to be magnetically active due to their strong magnetic fields generated by dynamos in thick convective envelopes \citep[e.g.,][]{Pallavicini1981ApJ248.279,Noyes1984ApJ279.763,Pizzolato2003AA397.147,Wright2011ApJ743.48,Reiners2014ApJ794.144}. 
They frequently produce superflares that appear as impulsive emission enhancements over a wide wavelength range. 
These superflares can each release a total bolometric energy of $\gtrsim 10^{33} \ \mathrm{erg}$ \citep[e.g.,][]{Chang2017ApJ834.92,Yang2017ApJ849.36,Gunther2020AJ159.60}, thus imposing severe impacts on the space environment and the habitability of nearby exoplanets \citep[e.g.,][]{Linsky2019LNP955,Varela2022SW20.e2022SW003164}. 

\indent In the past decades, stellar flares have received extensive statistical study 
at specific wavelength bands, especially optical \citep[e.g.,][]{Maehara2012Nat485.478,Shibayama2013ApJS209.5,Notsu2013ApJ771.127,Wu2015ApJ798.92,Maehara2015EPS67.59,Yang2017ApJ849.36,Tu2020ApJ890.46,Doyle2020MNRAS494.3596} 
and soft X-ray \citep[e.g.,][]{Pye2015AA581.A28,Tsuboi2016PASJ68.90,Getman2021ApJ916.32,Getman2021ApJ920.154,Zhao2024ApJ961.130} bands. 
An important finding is that they tend to show a significant statistical similarity with solar flares, independent of the observed wavelength band. 
The occurrence rate distribution against flare energy $\mathrm{d}N_\mathrm{flare} / \mathrm{d}E_\mathrm{flare} \propto E_\mathrm{flare}^{\gamma}$ at different bands shows a power-law form with a spectral index of $\gamma \sim -1.8$, for both solar and stellar flares. 
It implies that they may share the same energy release mechanism, most likely magnetic reconnection \citep[e.g.,][]{Shibata1995ApJL451.L83}, 
although stellar flares usually release energy greater than solar flares by several orders of magnitude. 
However, for a particular stellar flare, such speculation has not been justified by direct observations because simultaneous observations at as many wavelengths as possible are not easily accessible. 

\indent Impulsive radio enhancements known as radio bursts often accompany solar flares and have also been observed to exist on some magnetically active \citep[e.g.,][]{Crosley2018ApJ862.113,Villadsen2019ApJ871.214,Zic2020ApJ905.23,Zhang2023ApJ953.65,Zhang2024MNRAS531.919,Mohan2024AA686.A51} or even inactive \citep[e.g.,][]{Vedantham2020NA4.577,Callingham2021NA5.1233} M-dwarfs. 
\citet{Zhang2023ApJ953.65}, \citet{Zhang2024MNRAS531.919}, and \citet{Mohan2024AA686.A51} successively detected metric and decimetric radio bursts from a young and particularly active M-dwarf, AD Leo, and argued that the electron cyclotron maser (ECM) emission, one of the coherent emission processes, is a leading candidate mechanism to account for the observed signals. 
Alternatively, the radio signals detected from those magnetically inactive or quiescent stars could also originate from star-planet interactions \citep[e.g.,][]{Cauley2019NA3.1128,Vedantham2020NA4.577,Pope2021ApJL919.L10,Callingham2021NA5.1233,Pineda2023NA7.569}. 
For a majority of solar metric and decimetric (such as type-III and spike) radio bursts, the variation of their brightness temperature is observed to be almost synchronized in time with that of the hard X-ray and white-light emissions, thus providing strong evidence for electrons accelerated in processes associated with magnetic reconnection during flares \citep[e.g.,][]{Bastian1998ARAA36.131,Nindos2008SoPh253.3,Melrose2016SP291.3637,Benz2017LRSP14.2}. 
On the other hand, there are also multi-wavelength studies suggesting that solar decimetric spikes do not always originate from coronal X-ray flare sources \citep[e.g.,][]{Benz2002AA383.678,Battaglia2009AA499.L33}. 
Nevertheless, similar observations are very rare for stellar flares because, on the one hand, the brightness of comparable radio bursts from distant stars could be too weak to be detected; on the other hand, it would be extremely costly to monitor target stars for an extended period to wait for a flare. 

\indent In this paper, we report observations of groups of radio spikes that took place during a superflare on EQ Peg, using a combination of the Five-hundred-meter Aperture Spherical radio Telescope (FAST), which is the largest single-dish radio telescope and is sufficiently sensitive to search for weak signals from distant active stars, and the Transiting Exoplanet Survey Satellite (TESS). 
The high similarity of the finely structured radio bursts to solar ones and the simultaneity with the maximum of the white-light flare suggest that the radio spikes of interest originate from electrons accelerated by the superflare.

\section{Results}\label{section_results}
\subsection{Observations}\label{section_observation_results}

\indent Our target, EQ Peg (EQ Pegasi, also known as Gliese 896), is a binary system consisting of two M-dwarfs, and both components (EQ Peg A and B) are known to be flare-active stars \citep[e.g.,][]{Lacy1976ApJS30.85, Samus'2017AR61.80}. 
EQ Peg is one of the most extensively observed targets of magnetically active stars, surrounded by a giant exoplanet (EQ Peg Ab) with a mass of $2.26 \pm 0.57$ Jupiter masses and an orbital period of $284.39 \pm 1.47 \ \mathrm{days}$ \citep{Curiel2022AJ164.93}. 
Flaring activities on EQ Peg have been reported at radio \citep{Kundu1988AA195.159,Crosley2018ApJ856.39,Crosley2018ApJ862.113,Villadsen2019ApJ871.214}, optical \citep{Bromage1983ASSL102.245,Baliunas1984ApJ282.728,Katsova2002ASPC277.515,Dal2010AJ140.483,Dal2011AJ141.33,Kowalski2011ASPC448.1157,Dal2012NewA17.399,Crosley2018ApJ856.39}, ultraviolet \citep{Bromage1983ASSL102.245,Baliunas1984ApJ282.728}, and soft X-ray \citep{Poletto1986ASR6.145,Kundu1988AA195.159,Pallavicini1990AA228.403,Katsova2002ASPC277.515,Karmakar2022MNRAS509.3247} bands. 
Unfortunately, among these previous studies of EQ Peg, no radio bursts showing a strong association with an optical or soft X-ray flare have been observed. 

\indent TESS conducted optical and near-infrared ($6000 - 10000 \ \text{\AA}$) photometric observations of EQ Peg in Sector 56 for approximately 28 days (Sep. 2nd to 30th, 2022) with a cadence of 20 seconds. 
During the TESS observations, we observed EQ Peg with FAST at the L-band ($1.05 - 1.45 \ \mathrm{GHz}$), which was granted in the approved observation project \texttt{PT2022\_0136}. 
The observations were conducted during $16:16 - 18:46 \ \mathrm{UTC}$ on Sep. 8th, 2022 and during $16:18 - 18:48 \ \mathrm{UTC} $ on Sep. 9th, 2022, 
with a total of $5.0 \ \mathrm{h}$ tracking exposures. 

\begin{figure}
\centering
\includegraphics[width=0.5\textwidth]{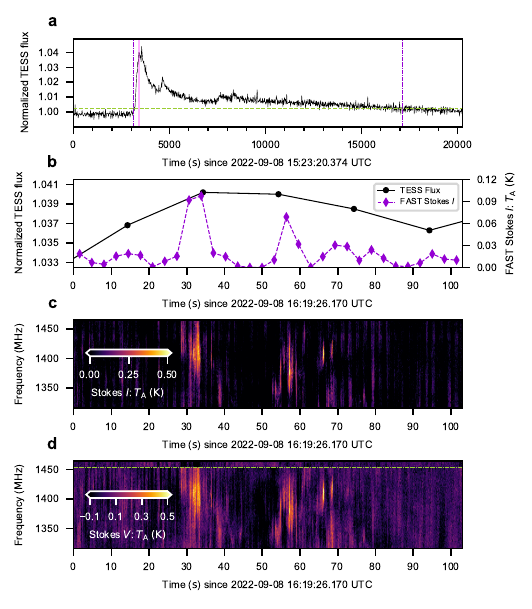}
\caption{TESS light curve and FAST dynamic spectra. 
(a) Light curve expressed in normalized TESS flux (the black solid line). 
The horizontal yellow-green dashed line shows the $3\sigma$ level above the quiescent flux ($1.00$), and the phase with flux exceeding $3\sigma$ is identified as the flaring phase. 
The start time and end time of the white-light flare are shown as vertical dark-violet dashed-dotted lines. 
The magenta rectangular shade indicates the time range when the most significant radio burst was detected by FAST, which is also the time range of panels (b), (c) and (d). 
(b) Zoom-in of the TESS light curve (black solid line with circles), 
and FAST Stokes $I$ rebinned to a time resolution of $3.2 \ \mathrm{s}$ (violet dashed line with diamonds), which is retrieved by averaging the dynamic spectrum in (c) against frequency channels. 
(c) FAST dynamic spectrum of Stokes $I$ showing the most significant radio burst we detected. 
(d) FAST dynamic spectrum of Stokes $V$. 
The horizontal yellow-green dashed-dotted line is the frequency of $1453 \ \mathrm{MHz}$, above which Stokes $V$ shows discontinuity in the frequency domain, and these data are excluded when calculating the degrees of circular polarization. 
The dynamic spectra of (c) and (d) are expressed in antenna temperature $T_\mathrm{A}$, and the colorbars are illustrated inside the corresponding panels. 
}
\label{TESS_lc+FAST_ds}
\end{figure}

\indent The TESS light curve and the FAST dynamic spectra are shown in Fig. \ref{TESS_lc+FAST_ds}, evidencing a white-light flare appearing in the period of $16:15:30 - 20:08:51 \ \mathrm{UTC}$ on Sep. 8th, 2022, with a duration of $3.9 \ \mathrm{h}$ (rise time of $0.1 \ \mathrm{h}$ and decay time of $3.8 \ \mathrm{h}$) and total bolometric energy of 
$3.6_{-1.0}^{+1.4} \times 10^{33} \ \mathrm{erg}$ for EQ Peg A or 
$1.2_{-0.3}^{+0.5} \times 10^{33} \ \mathrm{erg}$ for EQ Peg B, which is then classified as a superflare \citep[e.g.,][]{Maehara2012Nat485.478} for either component of this binary system (see Appendix \ref{section_methods} Methods for the details of the TESS data analysis). 
The TESS white-light flux increased by $\sim 4\%$ at the peak. 
Although the FAST exposure covered a large fraction of the white-light flare, 
the most significant radio burst appeared near the peak of the white-light flare and lasted for $\sim 100 \ \mathrm{s}$. 
Their temporal variations in the two distinct bands seem to show a correlation as illustrated in Fig. \ref{TESS_lc+FAST_ds}b. 
It is noted that, besides the most prominent one, other relatively weak radio bursts were also detected (see Fig. \ref{full_FAST_ds_modified} for the full FAST dynamic spectra). 

\indent The fine structures of the radio burst are well resolved, thanks to the extremely high sensitivity and temporal resolution of FAST. 
Totally $87$ local peaks with values greater than $5$ times the standard deviation are identified for this intense burst, 
as shown in Fig. \ref{FAST_ds}f-i (see Appendix \ref{section_methods} Methods for the details of the FAST data analysis). 
The properties of these local peaks are a good representation of spikes in the dynamic spectra. 
\begin{figure*}
\centering
\includegraphics[width=0.9\textwidth]{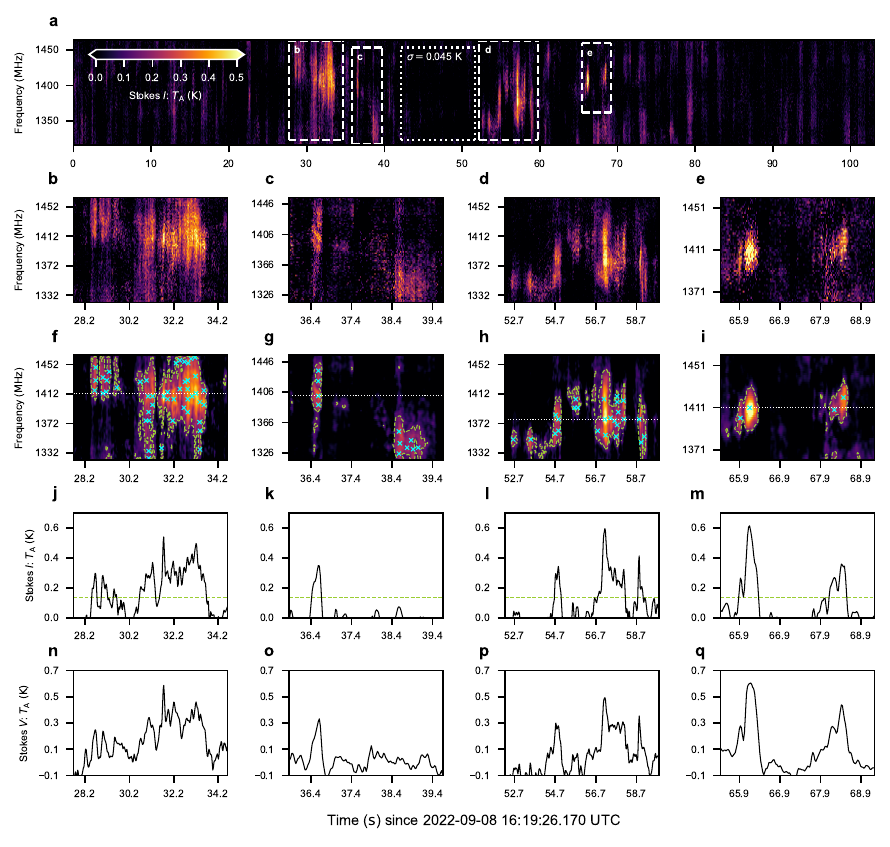}
\caption{Fine structures in the FAST dynamic spectra. 
(a) FAST dynamic spectrum of Stokes $I$. 
Axes and the colorbar are the same as those in Fig. \ref{TESS_lc+FAST_ds}c. 
The dynamic spectra are rebinned in both time and frequency in order to increase the signal-to-noise ratio, giving a time resolution of $25.16 \ \mathrm{ms}$ and a frequency resolution of $1.95 \ \mathrm{MHz}$. 
The four white dashed rectangular frames represent the time and frequency ranges of the dynamic spectra in (b)-(e), respectively. 
The white dotted rectangular frame represents the time and frequency ranges used to calculate the noise level, i.e. the standard deviation in the quiescence, giving 
$\sigma = 0.045 \ \mathrm{K}$ in the antenna temperature $T_\mathrm{A}$. 
(b)-(e) Zoom-in of the four sub-regions illustrated in (a). 
(f)-(i) Corresponding dynamic spectra with a Gaussian filter of (b)-(e). 
The yellow-green dashed contours show the $3\sigma$ level, and the cyan crosses denote the identified $87$ spikes with values greater than $5\sigma$. 
The horizontal white dotted lines indicate the central frequencies of the frequency channels used for retrieving the light curves in (j)-(m). 
(j)-(m) Light curves of Stokes $I$ retrieved from the frequency channels indicated in (f)-(i). 
The horizontal yellow-green dashed lines show the $3\sigma$ level. 
(n)-(q) Light curves of Stokes $V$, which are retrieved with the same method as (j)-(m), but from the dynamic spectra of Stokes $V$. 
}
\label{FAST_ds}
\end{figure*}

\begin{figure*}
\centering
\includegraphics[width=0.9\textwidth]{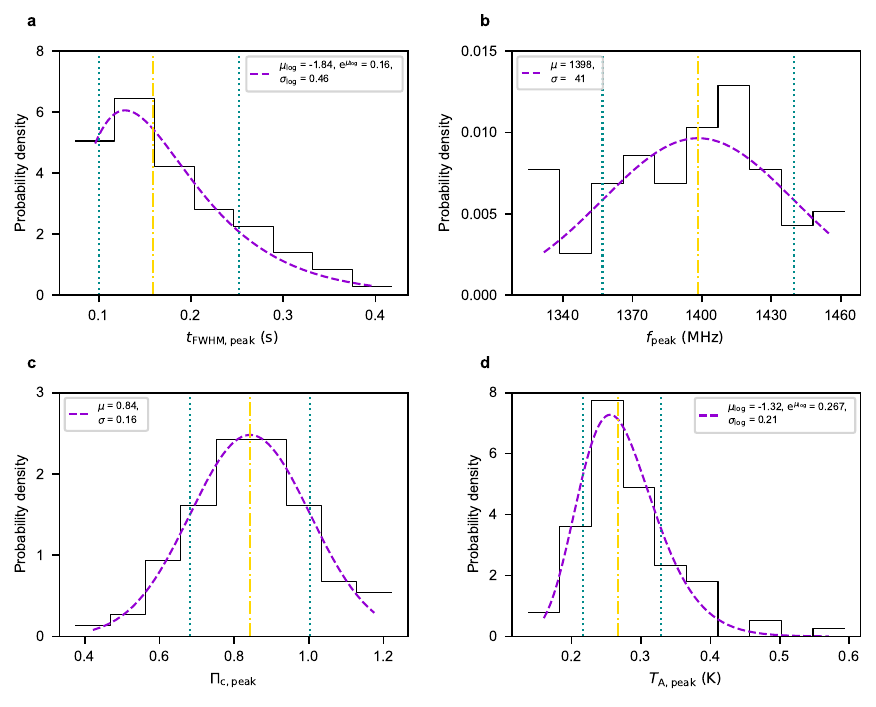}
\caption{Statistics of the most significant radio spikes in the FAST dynamic spectra. 
(a) Distribution of the FWHMs in the time domain $t_\mathrm{FWHM,peak}$. 
(b) Distribution of the frequencies $f_\mathrm{peak}$. 
(c) Distribution of the degrees of circular polarization $\mathit{\Pi}_\mathrm{c,peak}$. 
(d) Distribution of the antenna temperatures $T_\mathrm{A,peak}$. 
The histograms or probability density functions of $f_\mathrm{peak}$ and $\mathit{\Pi}_\mathrm{c,peak}$ are fitted by normal distributions, while those of $t_\mathrm{FWHM,peak}$ and $T_\mathrm{A,peak}$ are fitted by log-normal distributions. 
The best fits are illustrated with dark-violet dashed lines within each panel. 
As for the normal distributions, the mean ($\mu$) and the standard deviation ($\sigma$) of the best fit are listed in the panel legends; 
while as for the log-normal distributions, the mean ($\mu_\mathrm{log}$) and the standard deviation ($\sigma_\mathrm{log}$) of the best fit in the logarithmic scale are listed in the panel legends. 
As for the normal distributions, the values of $\mu$ are illustrated with yellow dashed-dotted lines, and the values of $\mu-\sigma$ and $\mu+\sigma$ are illustrated with cyan dotted lines; 
while as for the log-normal distributions, the values of $\mathrm{e}^{\mu_\mathrm{log}}$ are illustrated with yellow dashed-dotted lines, and the values of $\mathrm{e}^{\mu_\mathrm{log}-\sigma_\mathrm{log}}$ and $\mathrm{e}^{\mu_\mathrm{log}+\sigma_\mathrm{log}}$ are illustrated with cyan dotted lines. 
}
\label{statistics_peaks}
\end{figure*}

\indent We statistically analyze the properties of the spikes, including their 
full widths at half-maxima (FWHMs) in the time domain $t_\mathrm{FWHM,peak}$, 
frequencies $f_\mathrm{peak}$, 
degrees of circular polarization $\mathit{\Pi}_\mathrm{c,peak}$, 
and antenna temperatures $T_\mathrm{A,peak}$, as shown in Fig. \ref{statistics_peaks}. 
The probability density functions (PDFs) of $f_\mathrm{peak}$ and $\mathit{\Pi}_\mathrm{c,peak}$ are well fitted by normal distributions, while the PDFs of $t_\mathrm{FWHM,peak}$ and $T_\mathrm{A,peak}$ are well fitted by log-normal distributions. 
Note that we apply normal and log-normal distributions here only to statistically estimate the corresponding characteristic values and dispersions of the four parameters without specific physical assumptions. 
As for the normal distributions, we take the mean ($\mu$) and the standard deviation ($\sigma$) of the best fit as the characteristic value and dispersion, respectively. 
Regarding the log-normal distributions, we take the mean ($\mu_\mathrm{log}$) and the standard deviation ($\sigma_\mathrm{log}$) of the best fit in the logarithmic scale as the basis for calculating the characteristic value and dispersion, respectively. 
Therefore, the four parameters of the radio spikes are concentrated around characteristic values as 
\begin{equation} \label{local_peaks_distributions}
\begin{aligned}
t_\mathrm{FWHM,peak}         &\sim 0.159_{-0.059}^{+0.093} \ \mathrm{s}, \\
f_\mathrm{peak}              &\sim 1398 \pm 41 \ \mathrm{MHz}, \\
\mathit{\Pi}_\mathrm{c,peak} &\sim 0.84 \pm 0.16 , \\
T_\mathrm{A,peak}            &\sim 0.267_{-0.051}^{+0.062} \ \mathrm{K}. 
\end{aligned}
\end{equation}

\indent The FWHMs in the time domain represent the duration of each spike. 
Given the relatively limited bandwidth of our FAST observations ($\sim 0.5 \ \mathrm{GHz}$), the true distribution of $f_\mathrm{peak}$ of this radio-spike population is not necessarily centered around $1398 \ \mathrm{MHz}$. 
The fitted $f_\mathrm{peak}$ distribution is therefore adopted only as a reference for subsequent estimates of the magnetic field strengths in the radio source regions (see Section \ref{section_determination} for details). 
We adopt the IAU/IEEE convention \citep[e.g.,][]{Hamaker1996AAS117.161} which defines $\mathrm{Stokes} \ V$ as right-hand circularly polarized radiation minus left-hand circularly polarized radiation. 
Therefore, the positive values of $\mathit{\Pi}_\mathrm{c,peak}$ here indicate that the radiation of the spikes is right-hand circularly polarized. 
The noise level in antenna temperature is $0.045 \ \mathrm{K}$ (see Appendix \ref{section_methods} Methods for the details of the noise calculation), giving a signal-to-noise ratio (SNR) at $\sim 5.9$. 
Considering that the gain of the FAST central beam is $\sim 16 \ \mathrm{K \ Jy^{-1}}$ at $\sim 1400 \ \mathrm{MHz}$ within the zenith angle of $26.4^\circ$ \citep{Jiang2020RAA20.064J}, the observed flux densities of the spikes $F_\mathrm{peak}$ are estimated to be concentrated at $\sim 16.7_{-3.2}^{+3.9} \ \mathrm{mJy}$. 
By assuming the projected area of the superflare $A_\mathrm{flare}$ at peak time as an upper limit of the projected radio source area $A_\mathrm{RB}$, which gives 
$A_\mathrm{RB,A} \lesssim 1.2 \times 10^{-3} \pi R_{\star,\mathrm{A}}^2$ 
for EQ Peg A or 
$A_\mathrm{RB,B} \lesssim 8.1 \times 10^{-4} \pi R_{\star,\mathrm{B}}^2$ 
for EQ Peg B, 
the brightness temperatures of the observed spikes are estimated to be 
$T_\mathrm{B,peak} \gtrsim 4.5 \times 10^{13} \ \mathrm{K}$ 
for EQ Peg A or 
$T_\mathrm{B,peak} \gtrsim 1.3 \times 10^{14} \ \mathrm{K}$ 
for EQ Peg B 
(see Equation \ref{flare_area_Eq} and Equation \ref{brightness_temperature_Eq} in Appendix \ref{section_methods} Methods for the details of the estimation of the projected flare areas and the brightness temperatures, respectively). 
Here $R_{\star,\mathrm{A}}$ and $R_{\star,\mathrm{B}}$ are the stellar radii of EQ Peg A and B, which are $0.35$ and $0.25$ times the solar radius, respectively \citep{Morin2008MNRAS390.567}. 
The bandwidths $\Delta f$ of the spikes are estimated to be $\lesssim 50 \ \mathrm{MHz}$, which leads to $\Delta f / f_\mathrm{peak} \lesssim 0.04$; 
although some spikes exceed the upper limit of $\sim 1450 \ \mathrm{MHz}$. 

There are relatively weak bursts in the FAST dynamic spectra outside the peak time of the white-light superflare, as illustrated in Fig. \ref{full_FAST_ds_modified}. 
Similar properties can also be found for the weak signals, 
however, they are not included in the statistics because of their low SNRs (i.e., mostly $T_\mathrm{A} < 2\sigma$ in Stokes $I$).

\subsection{Emission mechanism}\label{section_emission_mechanism}

\indent The presence of extremely high brightness temperatures and high degrees of circular polarization for the finely structured bursts indicates that they are generated by a coherent emission process \citep[e.g.,][]{Dulk1985ARAA23.169}. 
The two most plausible coherent emission mechanisms are plasma and ECM emissions that emit at the plasma frequency $\nu_\mathrm{p}$ and the electron cyclotron frequency $\nu_B$, along with their harmonics, respectively. 
Their characteristic frequencies are directly correlated with the electron density $n_\mathrm{e}$ and the magnetic field strength $B$ as 
\begin{align}
\nu_\mathrm{p} \left[ \mathrm{Hz} \right] &\approx 8.98 \times 10^{3} \left( n_\mathrm{e} \left[ \mathrm{cm}^{-3} \right] \right)^{1/2}, \\
\nu_B \left[ \mathrm{Hz} \right]          &\approx 2.80 \times 10^{6} B \left[ \mathrm{G} \right], \label{nu_B}
\end{align}
respectively. 
ECM emission tends to occur when the local electron cyclotron frequency $\nu_B$ exceeds the local plasma frequency $\nu_\mathrm{p}$ at the emission location \citep[e.g.,][]{Melrose1982ApJ259.844,Dulk1985ARAA23.169}. 
This is guaranteed on M-dwarfs with strong magnetic fields, such as our target EQ Peg, 
since the average surface magnetic field $\left\langle B \right\rangle$ of EQ Peg A is observed to be $3.6_{-0.2}^{+0.4} \ \mathrm{kG}$ while that of EQ Peg B is about $4.2 \pm 1.0 \ \mathrm{kG}$ \citep{Shulyak2017NA1.184}. 
Besides, the ECM emission is more plausible for explaining the short-duration ($\lesssim 1 \ \mathrm{s}$) nature of the radio spikes due to the extraordinarily high maser growth rate, as argued by \citet{Treumann2006AAR13.229,Vedantham2021MNRAS500.3898,Zhang2023ApJ953.65}. 

\indent The ECM emission is commonly believed to be primarily in the extraordinary mode (x-mode) emission, and the consistent right-hand circular polarization  of the x-mode emission, as observed in this work, suggests that the radial component of the magnetic field is directed towards the observer \citep[e.g.,][]{Villadsen2019ApJ871.214}. 
That is to say, the emission we observed originates mostly from the northern magnetic hemisphere, if a large-scale dipolar magnetic field configuration is assumed. 
Previous Zeeman-Doppler imaging (ZDI) measurements of the magnetic fields also suggest that most of the visible part of the magnetic field corresponds to the northern magnetic hemisphere for both EQ Peg A and B \citep{Morin2008MNRAS390.567}, which is consistent with our inference of the direction of the magnetic field, although polarity reversals might occur during the time between their observations and ours. 
It is even reported that the reconstructed magnetic energy in the dipole mode accounts for approximately $70\%$ and $79\%$ of the total magnetic energy for EQ Peg A and B, respectively, and the orientation of the north poles is inclined at an angle of $60^\circ \pm 20^\circ$ with respect to the line of sight for both components \citep{Morin2008MNRAS390.567}.

\subsection{Determination of sources of radio spikes}\label{section_determination}
\indent EQ Peg was observed with all 19 beams of FAST \citep{Jiang2020RAA20.064J} simultaneously, aiming to inspect the influences of radio frequency interferences (RFIs) and the radio sources in the vicinity of EQ Peg. 
Signals with temporal characteristics similar to those of the central beam should be detected in other beams if radio bursts from nearby celestial bodies or RFIs occur. 
However, no such variable signals were detected in the 18 side beams, except for the central one, indicating negligible influences from RFIs or nearby radio sources. 

\indent Besides, we also inspected radio sky survey images (e.g., the Rapid ASKAP Continuum Survey, \citet{McConnell2020PASA37.e048}) around EQ Peg to search for bright background sources, but no sources with flux densities greater than or comparable to our target source were found within the $\sim 3 \ \mathrm{arcmin}$ half-power beamwidth of the FAST receiver at $\sim 1400 \ \mathrm{MHz}$ \citep{Jiang2020RAA20.064J}. 
This reconfirms that the radio spikes are from our target source instead of RFIs or background radio sources. 

\indent The possibility that the radio spikes originate from the interactions of the magnetic fields of the two component stars can also be excluded. 
RS CVn-type binaries are a common type of radio-active binaries with magnetic field interactions \citep[e.g.,][]{Uchida1983ASSL102.629}. 
These binaries are close binaries and have typical orbital periods of $\lesssim 100 \ \mathrm{days}$ \citep{Hall1976IAUC60.287}. 
In comparison, the orbital period of EQ Peg is $8.37 \times 10^4 \ \mathrm{days}$, with a semi-major axis of $31.6 \ \mathrm{AU}$ \citep{Curiel2022AJ164.93}. 
The magnetic field strength near the midpoint of the two components is $\sim 10^{-10} \ \mathrm{G}$ in magnitude, making them distant enough to exclude the possibility of being an interacting close binary system. 
Given that the detected radio bursts have typical characteristics, including durations, polarizations, and brightness temperatures, akin to solar radio spike bursts \citep[e.g.,][]{Dulk1985ARAA23.169,Nindos2008SoPh253.3}, as well as a high correlation with the maximum of the white-light emission, they are confidently concluded to originate from the superflare on one of the components, i.e., EQ Peg A or B. 
Note that, as neither FAST nor TESS can spatially resolve the two components, both situations will be discussed in the following. 

\indent As suggested by Equation (\ref{nu_B}), the frequency of the ECM emission is directly related to the local magnetic field strength, which can thus restrict the emission location from observed spikes once the magnetic field configuration is known. 
Usually, the fundamental radiation has the highest growth rate for the cyclotron maser instability, and the growth rate decreases as the order of the harmonic increases, meaning that the fundamental radiation is the most significant \citep[e.g.,][]{Wu1985SSR41.215,Aschwanden1990AA237.512}. 
The frequency range of 
$1398 \pm 41 \ \mathrm{MHz}$ 
as in Equation (\ref{local_peaks_distributions}) corresponds to a range of magnetic field strengths of 
$485 - 515 \ \mathrm{G}$ 
if the fundamental radiation is assumed. 

\indent Previous studies utilizing the ZDI method have shown that both EQ Peg A and B have dipole-dominated magnetic fields, and the magnetic dipoles of the two have the same orientation relative to the line of sight \citep[e.g.,][]{Morin2008MNRAS390.567}. 
Assuming that the magnetic field configuration is a dipole for both components, the field strength distribution $B (r,\theta,\varphi)$ is then denoted as 
\begin{equation}
B (r,\theta,\varphi) = \frac{B_\mathrm{polar}}{2} \frac{R_\star^3}{r^3} \sqrt{1 + 3 \cos^2{\theta}} 
\end{equation}
in the spherical coordinate system, 
where $B_\mathrm{polar}$ is the magnetic field strength at the polar surface, $r$ is the radial distance from the center of the star, $\theta$ is the polar angle, and the polar direction closer to the line-of-sight direction (i.e., the magnetic north pole, as discussed in \ref{section_emission_mechanism} Emission mechanism) is set as $\theta=0$. 
$\varphi$ is the azimuthal angle, and the azimuthal angle of the line-of-sight direction is set as $\varphi=0$. 
Very recently, it has been demonstrated that, for such a large-scale dipolar configuration, the rapid rotation, for example, of our target EQ Peg can facilitate the generation of plasmoid eruptions and high-lying, high-temperature loops across the equator \citep{Daley-Yates2024MNRAS534.621}. 
It suggests the plausibility of large-scale flare loops situated above the equator and embedded in a global dipolar field. 

Here, we mainly adopt the magnetic fields measured by a recent study \citep{Shulyak2017NA1.184}, which gives $B_\mathrm{polar,A} = 5.2_{-0.3}^{+0.6} \ \mathrm{kG}$ for EQ Peg A and $B_\mathrm{polar,B} = 6.1 \pm 1.4 \ \mathrm{kG}$ for EQ Peg B. 
The possible emission locations are then restricted to a shell surrounding the star, and this shell is lower at the equator ($\theta = 90^\circ$) and higher at the poles ($\theta = 0^\circ$ and $180^\circ$). 

\indent Loss-cone or horseshoe velocity distributions of mildly relativistic electrons are prerequisites to generating the ECM \citep{Treumann2006AAR13.229,Bingham2013SSR178.695}. 
Loss-cone-driven ECM emission has been studied in solar \citep[e.g.,][]{Aschwanden1988ApJ332.447,Aschwanden1990AAS85.1141} and stellar contexts \citep[e.g.,][]{Zarka2025AA695.A95}, 
which provide theoretical supports for this mechanism in solar and stellar coronal loops. 
Loss-cone-distributed electrons generate ECM radiation with directions oblique to the magnetic field lines, 
and the beaming angle $\alpha$ at a certain location, i.e., the emission direction relative to the local magnetic field line, can be expressed as 
\begin{equation} \label{loss-cone_beaming_angle_eq}
\begin{aligned}
&\alpha(r,\theta,\varphi) =\\ 
&\arccos\left[{\frac{v/c}{\sqrt{1-\nu_B(r,\theta,\varphi)/\nu_{B,\mathrm{max}}(r,\theta,\varphi)}}}\right] 
\end{aligned}
\end{equation}
based on a simplified loss-cone-driven ECM model \citep[e.g.,][]{Hess2008GRL35.L13107,Louis2019AA627.A30,Kavanagh2023MNRAS524.6267}, 
where $v$ is the electron velocity, $c$ is the speed of light, and $\nu_{B,\mathrm{max}}$ is the maximum electron cyclotron frequency that the electrons can reach along the field line, i.e. at the field line footpoint. 
Spatial distributions of the beaming angle for ECM driven by loss-cone-distributed electrons with kinetic energies of $5,\ 10,\ 20 \ \mathrm{keV}$ are illustrated by Fig. \ref{loss-cone_beaming_angle}. 

\indent Horseshoe-distributed electrons have been largely proven to exist in the solar flare loops \citep[e.g.,][]{Melrose2016SP291.3637,Ning2021AA651.A118}. 
The emission regions associated with horseshoe-distributed electrons can be constrained with fewer free parameters than those of loss-cone-distributed electrons, since they usually generate ECM radiation with directions quasi-perpendicular to the magnetic field lines. 
The ECM emission driven by both distributions are highly beamed, 
forming a radiation cone with the axis in a certain oblique direction given by Equation (\ref{loss-cone_beaming_angle_eq}) or perpendicular to the field line, 
and the apex angle of only several degrees \citep[e.g.,][]{Treumann2006AAR13.229,Bingham2013SSR178.695}. 

\begin{figure*}
\centering
\includegraphics[width=0.8\textwidth]{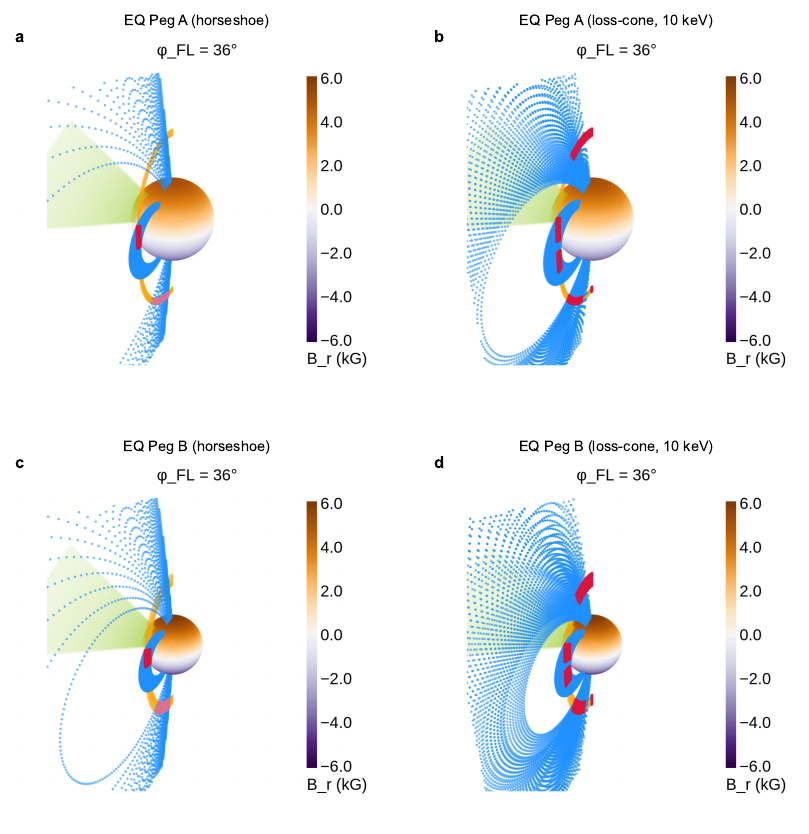}
\caption{Tentative determination of sources of the radio bursts. 
(a) A 3D illustration of the radio sources as for EQ Peg A under the condition of horseshoe-distributed electrons. 
The yellow-green fan-shaped area indicates the line-of-sight directions or inclination angles ($i = 60^\circ \pm 20^\circ$). 
The sphere suggests the scale of EQ Peg A and the colormap denotes the radial component of the magnetic field $B_r$ on the surface, where only the central value corresponding to $\left\langle B \right\rangle = 3.6 \ \mathrm{kG}$ is adopted. 
Positive values of $B_r$ suggest the northern magnetic hemisphere, while negative values suggest the southern magnetic hemisphere. 
The orange semi-ring denotes the locations where the magnetic field strength corresponds to the frequency range of $1398 \pm 41 \ \mathrm{MHz}$. 
The red regions on the orange semi-ring denote the emission sources after considering the direction of emission, 
while the darker one refers to the most possible source (see Section \ref{section_determination} Determination of sources of radio spikes for details), 
and the blue dots depict the field lines passing through the sources. 
(b) Similar to (a), but under the condition of loss-cone-distributed electrons with $10 \ \mathrm{keV}$ kinetic energy. 
Only the radio sources with $\theta$ between $10^\circ$ and $170^\circ$ are illustrated with corresponding magnetic field lines. 
(c) and (d) Similar to (a) and (b), but for the situations of EQ Peg B. 
The panels illustrate only the special situation where the azimuthal angle of the field lines ($\varphi_\mathrm{FL}$) is $36^\circ$ (see Section \ref{section_determination} Determination of sources of radio spikes for the detailed definition of this spherical coordinate system), 
and the snapshots showing more cases for different $\varphi_\mathrm{FL}$ are given in Fig. \ref{dipole_3d_snapshots_appendix_horseshoe} and Fig. \ref{dipole_3d_snapshots_appendix_loss-cone}. 
}
\label{dipole_3d}
\end{figure*}

\indent The magnetic field lines can be expressed simply as 
\begin{equation}
r = r_\mathrm{FL,equator} \sin^2{\theta} 
\end{equation}
under the assumption of a dipole configuration, where $r_\mathrm{FL,equator}$ is the distance between the top of the field line (i.e. the location above the equator of the star on the field line) and the center of the star. 
The $\theta$ of the line-of-sight direction (or the inclination angle $i$) for both EQ Peg A and B is $60^\circ \pm 20^\circ$ \citep{Morin2008MNRAS390.567}, which is depicted by the yellow-green fan-shaped areas in Fig. \ref{dipole_3d}. 
By calculating the emission directions at different locations on the field lines with azimuthal angles $\varphi_\mathrm{FL}$ ranging from $0^\circ$ to $180^\circ$ under horseshoe and loss-cone distributions, 
and inspecting whether they point to the line-of-sight direction, the possible emission locations are further restricted to several segments of each magnetic field line, as illustrated in Fig. \ref{dipole_3d} for both EQ Peg A and B. 
The horseshoe-driven emission sources are determined to concentrate on two small segments, while the loss-cone-driven ones tend to extend to wider areas. 

\indent Furthermore, based on the magnetic reconnection theory, \citet{Namekata2017ApJ851.91} proposed a semi-empirical method to derive the length scale $L_\mathrm{loop}$ of the solar and stellar flare loops from optical observations : 
\begin{equation}
\begin{aligned}
& L_\mathrm{loop} \sim \\ 
& 1.64 \times 10^{9} 
\left( \frac{\tau_\mathrm{decay}}{100 \ \left[ \mathrm{s} \right]} \right)^{2/5} 
\left( \frac{E_\mathrm{flare}}{10^{30} \ \left[ \mathrm{erg} \right]} \right)^{1/5} 
\ \left[ \mathrm{cm} \right], 
\end{aligned}
\end{equation}
where $\tau_\mathrm{decay}$ is the e-folding decay time, and $E_\mathrm{flare}$ is the flare energy. 
By using the observed decay time (see Appendix \ref{section_methods} Methods for details), i.e. $3.8 \ \mathrm{h}$, as an estimation of the e-folding decay time, we derive the length scale of the flare loop to be $\sim 2.5 R_\mathrm{\star,A}$ for EQ Peg A or $\sim 2.8 R_\mathrm{\star,B}$ for EQ Peg B. 
Besides, based on soft X-ray observations, \citet{Karmakar2022MNRAS509.3247} derived that the flare loop height is $\lesssim 5$ times the stellar radius of either EQ Peg A or B for superflares with soft X-ray energy $E_\mathrm{flare,SXR} \sim 10^{34} \ \mathrm{erg}$. 
Therefore, among the two segments under the horseshoe distribution, the one closer to the equator of the star is considered the most likely source, as highlighted by the darker colors in Fig. \ref{dipole_3d}a and c, and Fig. \ref{dipole_3d_statistics_appendix}. 
And for loss-cone-driven radio sources, the ones close to the equator are more likely than those close to the polar regions. 
Two animations (Fig. \ref{video_s1} and Fig. \ref{video_s2}) including all cases for $\varphi_\mathrm{FL}$ ranging from $0^\circ$ to $180^\circ$ are also available online. 
%

\indent Finally, assuming that the emission is confined within a solid angle of $10$ to $100$ square degrees, which is inferred from the apex angle (several degrees) of the radiation cone, and adopting a typical spike bandwidth of $\sim 50 \ \mathrm{MHz}$, the emitted radio power, or luminosity, of the spikes is estimated to range from $\sim 9.1 \times 10^{18}-1.2 \times 10^{20} \ \mathrm{erg \ s^{-1}}$.

\section{Summary and discussion}\label{section_discussion}
\indent We performed a multi-wavelength observation of EQ Peg at both radio (FAST) and optical (TESS) bands and detected groups of decimetric radio bursts near the peak of a white-light superflare. 
Thanks to the extremely high sensitivity and temporal resolution of FAST, the fine structures of the radio bursts, consisting of spikes, are clearly resolved. 
The radio spikes are argued to be generated through the ECM mechanism, given their extremely high brightness temperatures, high degrees of circular polarization, and extremely short durations. 
The co-occurrence of the finely structured radio bursts with the peak of the white-light superflare represents a strong correlation between a stellar radio burst and a superflare ever observed, supporting that they could be generated by the same physical process as solar flares, i.e., most likely by electrons accelerated in processes associated with magnetic reconnection. 

\indent Solar radio bursts often show frequency drifts in the dynamic spectra due to the movement of the radiation sources \citep[e.g.,][]{Dulk1985ARAA23.169}. 
However, in our observation of EQ Peg, the spikes occurred almost simultaneously within the $\Delta f \lesssim 50 \ \mathrm{MHz}$ bandwidth. 
No significant frequency drifts within the FAST dynamic spectra suggest that the emission originated almost simultaneously from flare loops with magnetic fields of a range of strengths, according to the correlation between the ECM emission frequency and the magnetic field strength. 
One possibility is that the trapped electrons conforming to the ECM mechanism indeed distribute over a relatively large range of altitudes, where the magnetic field has strengths of 
$485 - 515 \ \mathrm{G}$ 
and is approximately illustrated in Fig. \ref{dipole_3d}a and b. 
The emission generated by such a group of electrons at different altitudes is detected by the FAST almost simultaneously. 

\indent Another possibility for this zero-drift is that the electrons generating the observed spikes move along the flare loops at very large velocities, leading to very high frequency drift rates, which are hardly distinguishable within the relatively narrow ($\lesssim 50 \ \mathrm{MHz}$) bandwidth at the current time and frequency resolutions. 
Our Gaussian filter kernel has a radius of $4$ pixels in the smoothed dynamic spectra. 
In order to effectively distinguish the frequency drift, the spike should drift more than $\sim 4$ pixels in time within the $\sim 50 \ \mathrm{MHz}$ bandwidth, resulting in an upper limit of the distinguishable drift rate of $\sim 497 \ \mathrm{MHz \ s^{-1}}$. 
We further estimate the corresponding upper limits of the electron dynamic energy to be $\sim 7.3 \ \mathrm{keV}$ for EQ Peg A and $\sim 4.1 \ \mathrm{keV}$ for EQ Peg B, if only the radial velocity is considered. 
These values fall within the typical energy range required for ECM emission \citep[e.g.,][]{Treumann2006AAR13.229}, indicating that the actual electron energies could exceed these upper limits. 
In that case, the frequency drift will be indistinguishable at our current capabilities. 

\indent Nevertheless, according to the large length scale of the flare loop derived in Section \ref{section_determination}, as well as the consistency that all detected radio spikes exhibit right-hand circular polarization, we argue a source region highly possible in the large-scale dipolar field, although the spikes could be fast frequency-drifting. 

\indent Compared to single-band observations, simultaneous multi-wavelength observations here determine the origin of superflares more confidently \citep[also see,][]{Davis2025ApJ993.82}. 
Recent studies have also reported radio bursts from nearby magnetically active M-dwarfs. 
\citet{Zhang2023ApJ953.65} identified stripe-like and blob-like fine structures among the radio bursts from AD Leo and also suggested the mechanism of ECM. 
However, due to the lack of a correlation with an optical flare, it was not fully determined whether the radio bursts were from accelerated electrons associated with a flare or from a magnetospheric process such as co-rotation breakdown \citep{Zarka2025AA695.A95}. 
Moreover, \citet{Callingham2025Nat647.603} reported a stellar radio burst with characteristics consistent with a type-II burst, and interpreted it as evidence for a stellar coronal mass ejection. 
However, the lack of simultaneous multi-wavelength flare observations prevents a more comprehensive investigation of the associated energy release process. 
Similar uncertainties also exist in some earlier studies of radio burst observations from magnetically active stars \citep[e.g.,][]{Kundu1988AA195.159,Osten2006ApJ637.1016,Osten2008ApJ674.1078,Crosley2018ApJ862.113,Villadsen2019ApJ871.214}. 
Interestingly, \citet{Zic2020ApJ905.23} reported a possible solar-like decimetric radio burst $\sim 42 \ \mathrm{s}$ before the onset of an optical flare from Proxima Centauri. 
But the fine structures were not fully resolved due to the limited temporal and frequency resolutions; see also other similar observations \citep{Osten2005ApJ621.398,MacGregor2021ApJ911.L25,Tristan2025ApJ986.53}. 
In comparison, the radio bursts reported here not only exhibit a close temporal correlation to the white-light superflare but also present well-resolved fine structures with a high degree of circular polarization. 
It thus provides evidence for well-resolved solar-like spike bursts originating from flares on a distant star for the first time. 
It is encouraged to perform more FAST observations of active stars in the future to diagnose the energy release processes of various stellar flares and even estimate their influences on nearby exoplanets. 

\begin{acknowledgments}

\indent 
We thank Jiale Zhang (Peking University) for the helpful suggestions and comments on the FAST data analysis. 
This work is supported by the National Natural Science Foundation of China under grant 12525305, the Strategic Priority Research Program of the Chinese Academy of Sciences, Grant No. XDB0560000, and the Fundamental Research Funds for the Central Universities under grant 2025300318. 
This work has made use of the data from FAST (Five-hundred-meter Aperture Spherical radio Telescope; \url{https://cstr.cn/31116.02.FAST}). 
FAST is a Chinese national mega-science facility operated by the National Astronomical Observatories, Chinese Academy of Sciences. 
This paper includes data collected with the TESS mission, obtained from the MAST data archive at the STScI. 
Funding for the TESS mission is provided by the NASA Explorer Program. 
STScI is operated by the Association of Universities for Research in Astronomy, Inc., under NASA contract NAS 5-26555. 
The numerical calculation was done on the computing facilities in the High Performance Computing Center of Nanjing University. 

\end{acknowledgments}

\facilities{FAST:500m, TESS}

\appendix

\section{Supplementary discussions}
\subsection{Reliability of emission mechanism}
\indent In the standard model of solar flares, magnetic reconnection can accelerate electrons that stream down along flare loops and deposit energy to the photosphere through thermalization, giving rise to the white-light emission enhancement in the visible band \citep[e.g.,][]{Najita1970SP15.176,Aboudarham1986AA156.73}. 
At the same time, because part of the accelerated electrons are trapped within flare loops, the ECM emission will be generated once the velocity of trapped electrons presents a loss-cone or horseshoe distribution \citep[e.g.,][]{Treumann2006AAR13.229,Bingham2013SSR178.695}. 

\indent Besides the reconnection mechanism, alternative mechanisms may also contribute to the observed white-light emission. 
For instance, \citet{Heinzel2018ApJ859.143} proposed that flare loops, particularly under high-density conditions, as expected in active M dwarfs, can produce significant white-light emission through thermal processes and radiative backwarming during stellar superflares. 
Nevertheless, the close temporal correlation between the radio spike bursts that explicitly originate from non-thermal electrons and the white-light superflare in our observations is more inclined to support the scenario that both phenomena are powered by a common population of non-thermal electrons, 
which are most likely accelerated in processes associated with magnetic reconnection \citep{Shibata1999ApJ526.L49,Shibata2002ApJ577.422,Namekata2017ApJ851.91}. 

\indent The white-light emission of a solar flare usually coincides with the hard X-ray emission, which is another indicator of non-thermal electrons \citep[e.g.,][]{Kuhar2016ApJ816.6}, even though a temporal offset occasionally exists \citep[e.g.,][]{Namekata2017ApJ851.91}. 
Similarly, in our observations, the most significant radio burst occurred within the time window of the most intense white-light emission. The good synchronization in time thereby reinforces the interpretation of common non-thermal electrons.

\subsection{Uncertainty of emission sources}

\indent We also determine the possible emission locations of the radio spikes, including the heights and latitudes in the coronae of either EQ Peg A or B based on the ECM mechanism driven by horseshoe- and loss-cone-distributed electrons and the dipole field assumption. 
The similar conclusions were also obtained by \citet{Zarka2025AA695.A95} for circularly polarized radio bursts from another star, AD Leo, recently; however, they were not known to be associated with a superflare. 

\indent It is worth noting that our dipole field assumption is based on ZDI magnetic field measurements of EQ Peg, which usually underestimate the complexity of the magnetic field. 
In order to determine the sources of the detected radio spike bursts more precisely, high-resolution magnetograms of EQ Peg are thus highly desired in the future.

\subsection{Extremely short duration}

\indent The duration of each spike is extremely short, i.e., $t_\mathrm{FWHM,peak} < 1 \ \mathrm{s}$, which may be due to the extraordinarily high maser growth rate in the ECM emission process \citep[e.g.,][]{Treumann2006AAR13.229}. 

\indent On the other hand, the more stimulating aspect is that 
the total duration of the group of spikes is also short, i.e. $t_\mathrm{group,RB} \sim 800 \ \mathrm{s}$, which seems to be incompatible with the $3.9 \ \mathrm{h}$ duration of the accompanying white-light flare. We suggest the following two possibilities. 
First of all, the ECM emission only appears within a small angle range, which therefore requires a very specific direction to be observed at each moment. 
Since the FAST was in tracking mode, the line-of-sight direction did not change during the total $2.5 \ \mathrm{h}$ exposure. 
However, the radiation direction could always be varying, making the emitting radio signal only visible for a very short duration. 
One major reason for the variation in the direction of radiation is the rotation of the target stars. 
Both EQ Peg A and B have very short rotation periods, which are $P_\mathrm{rot,A} = 1.061 \pm 0.004 \ \mathrm{d}$ and $P_\mathrm{rot,B} = 0.404 \pm 0.004 \ \mathrm{d}$, respectively \citep{Morin2008MNRAS390.567}. 
Assuming rigid body rotation, the values of the corresponding angular velocities are 
$\omega_\mathrm{rot,A} = 0.0039 ^\circ \ \mathrm{s}^{-1}$ 
and 
$\omega_\mathrm{rot,B} = 0.0103 ^\circ \ \mathrm{s}^{-1}$, 
respectively. 
During a timescale of $\sim 800 \ \mathrm{s}$, they could rotate by 
$\Delta \varphi_\mathrm{rot,A} \sim 3.1^\circ$ 
and 
$\Delta \varphi_\mathrm{rot,B} \sim 8.3^\circ$, 
respectively. 
Assuming the radio source is roughly circular and applying the estimated radio source area, 
the corresponding linear scale or diameter $d_\mathrm{RB}$ is estimated to be 
\begin{equation}
d_\mathrm{RB} \sim 2 \sqrt{\frac{A_\mathrm{RB}}{\pi}}, 
\end{equation}
and the angular scale relative to the center of the star is then estimated to be 
\begin{equation}
\Delta \varphi_{\mathrm{RB}} \sim \frac{d_\mathrm{RB}}{R_\star} = \frac{2}{R_\star} \sqrt{\frac{A_\mathrm{RB}}{\pi}}. 
\end{equation}
The value of $\Delta \varphi_\mathrm{RB}$ is 
$\Delta \varphi_\mathrm{RB,A} \lesssim 4.0^\circ$ for EQ Peg A and 
$\Delta \varphi_\mathrm{RB,B} \lesssim 3.3^\circ$ for EQ Peg B, respectively. 
As discussed, 
the apex angle of the emission directions, or the emission cone, is also usually several degrees. 
The values of $\Delta \varphi_\mathrm{rot}$, $\Delta \varphi_\mathrm{RB}$, and the apex angle of the emission cone are all of the same order of magnitude, which suggests that the fast rotation of the target stars could account for the very short duration of the radio burst group, i.e., $t_\mathrm{group,RB} \sim 800 \ \mathrm{s}$. 
In contrast, such a fast rotation does not influence the visibility of the white-light emission as long as the corresponding source is on the disk. 

\indent Another possibility accounting for the short appearance of the group of radio spikes is that the energy release of the flare is usually most efficient at the peak time; 
therefore, the radio emissions are expected to be the strongest. 
While the radio emissions are relatively weaker, except during peak times, they are no longer detectable by FAST.

\subsection{Detection probability of radio bursts associated with superflares}

\begin{figure*}
\centering
\includegraphics[width=0.9\textwidth]{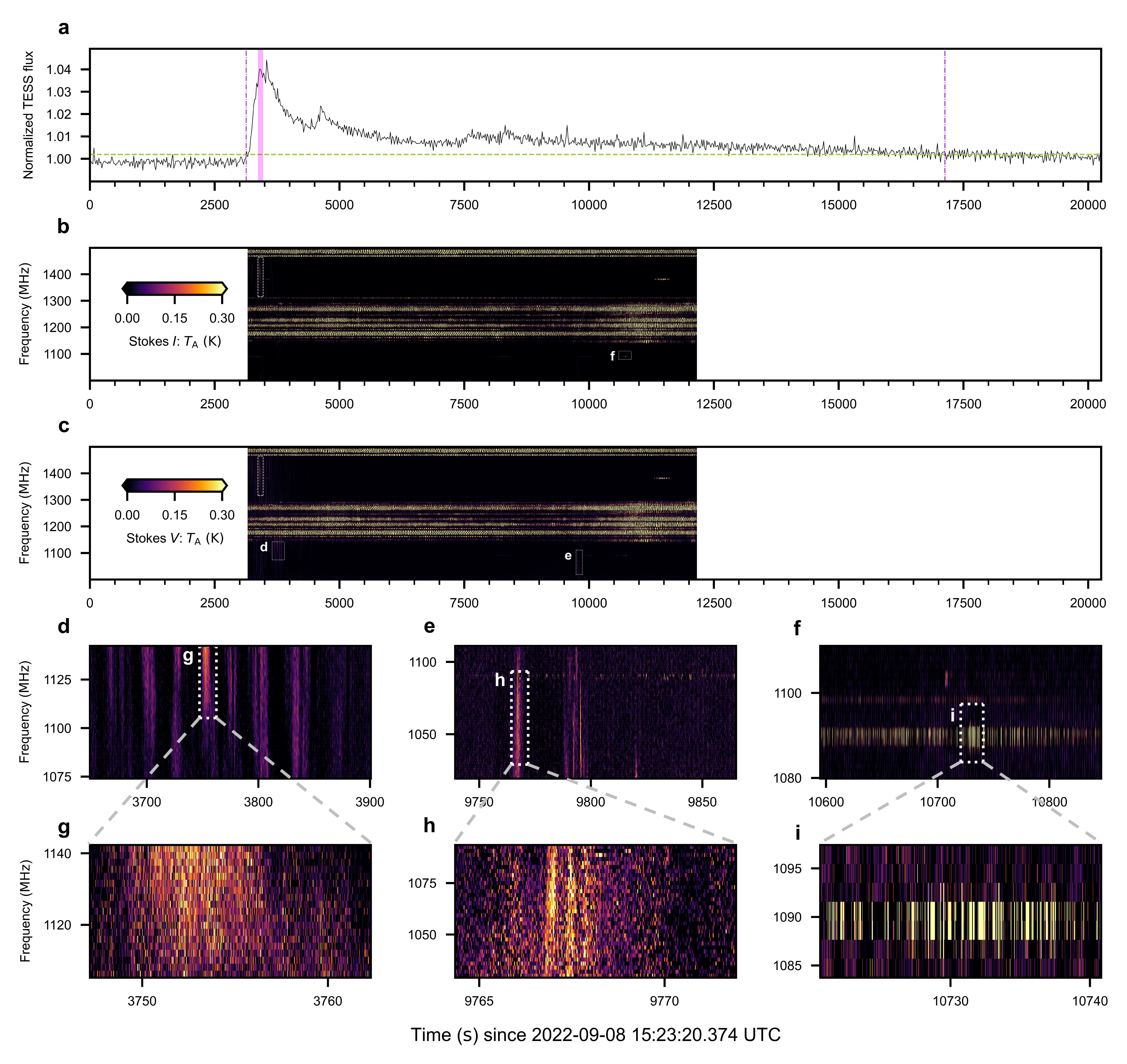}
\caption{(a) TESS light curve expressed in normalized TESS flux. 
The meanings of all the elements are the same as those in Fig. \ref{TESS_lc+FAST_ds}a. 
(b) FAST dynamic spectrum of Stokes $I$ during the entire $2.5 \ \mathrm{h}$ exposure on Sep. 8th, 2022. 
The white dashed rectangular frame represents the time and frequency ranges of the dynamic spectrum in Fig. \ref{TESS_lc+FAST_ds}c, among which the most significant signals were detected. 
(c) Similar to (b), but for Stokes $V$. 
The horizontal features at $\sim 1200 \ \mathrm{MHz}$ and $\sim 1500 \ \mathrm{MHz}$ in (b) and (c) are caused by RFIs. 
The three white dotted rectangular frames in (b) and (c) represent the time and frequency ranges of the dynamic spectra in (d)-(f), respectively. 
(g)-(i) Zoom-in of the corresponding sub-regions denoted in (d)-(f). 
The signals in (f) and (i) are most likely from RFIs rather than our targets. 
}
\label{full_FAST_ds_modified}
\end{figure*}

\indent During the 2.5-hour continuous observation by FAST, only several groups of radio bursts were detected; no other intense radio activities were observed (see Fig. \ref{full_FAST_ds_modified}). 
According to statistics of white-light flares on EQ Peg and the duration-energy correlation of flares on magnetically active mid M-dwarfs \citep[e.g.,][]{Lacy1976ApJS30.85,Maehara2021PASJ73.44}, it is estimated that the fraction of flares on EQ Peg with bolometric energy $\gtrsim 10^{32} \ \mathrm{erg}$ is only $\sim 0.007$. 
This shows that the possibility of radio bursts being randomly aligned with a white-light superflare is very low, 
especially for the one observed at the peak of the superflare. 
On the other hand, it was reported that for late M-dwarfs, there is not a strong correlation between a high flaring rate and its detectability at the radio bands, suggesting a potential decrease in ``radio efficiency'' of optical flares for later-type stars \citep{Yiu2024AA684.A3}. 
Nevertheless, the radio bursts occurring near the peak of the superflare were indeed observed for our target. 
The occurrence of this unlikely event indicates that they are most likely correlated in physics. 
However, to further confirm such a correlation, long-term monitoring of EQ Peg in white-light and radio bands, as well as conducting statistical studies, is essential.

\section{Methods}\label{section_methods}
\subsection{TESS data analysis}\label{section_TESS_data_analysis}
\begin{figure}
\centering
\includegraphics[width=0.5\textwidth]{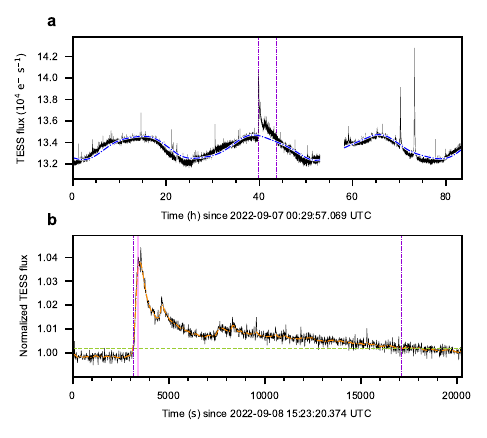}
\caption{Analysis of the TESS light curve. 
(a) A section of the original TESS light curve (the black solid line). 
The blue dashed-dotted line stands for the long-term trend of the stellar rotational modulation fitted by a periodic function. 
The gap at about $55 \ \mathrm{h}$ indicates no observations. 
The vertical dark-violet dashed-dotted lines suggest the identified start and end times of this white-light flare and represent the same time points as what the vertical dark-violet dashed-dotted lines illustrate in (b). 
(b) Normalized light curve (the black solid line). 
The orange dashed-dotted line illustrates the light curve smoothed by a Gaussian filter and is directly used in the identification of the start and end times. 
The meanings of all the other elements in this panel are the same as those in Fig. \ref{TESS_lc+FAST_ds}a. 
}
\label{TESS_lc_method}
\end{figure}

\indent We use the 20-second cadence Target Pixel File from the Mikulski Archive for Space Telescopes (MAST) at the Space Telescope Science Institute (STScI)\footnote{\url{https://mast.stsci.edu/}} 
and extract the light curve $F_\mathrm{TESS}(t)$ using the pipeline-defined aperture mask. 
These procedures, including searching for and downloading the data, and retrieving the TESS light curve, are performed using the Python package Lightkurve\footnote{\url{https://docs.lightkurve.org/}}, which is a dedicated package for analyzing light curves in time-domain astronomy. 
We obtain the long-term trend of the stellar rotational modulation by using a periodic function described in Equation (\ref{periodic_function}) 
\begin{figure*}
\centering
\begin{equation} \label{periodic_function}
F_\mathrm{TESS,trend}(t) = 
C + 
C_\mathrm{A} \sin{\left( \frac{2\pi}{P_\mathrm{rot,A}}t + \phi_\mathrm{A} \right)} + 
C_\mathrm{B} \sin{\left( \frac{2\pi}{P_\mathrm{rot,B}}t + \phi_\mathrm{B} \right)}. 
\end{equation}
\end{figure*}
to fit the entire light curve of EQ Peg during the 28-day observation in TESS Sector 56\footnote{\url{https://tess.mit.edu/observations/sector-56/}}, after removing the data points whose flux values are outlying the overall median value by at least 3 times the overall standard deviation. 
Most of the flares are removed from the light curve, and the main cause of the remaining flux variation is the rotational modulation of the stars after the removal of such outlying data points. 
Within this periodic function, $P_\mathrm{rot,A}$ and $P_\mathrm{rot,B}$ are the corresponding rotation periods of EQ Peg A and B, set as fixed values of $1.061 \ \mathrm{d}$ and $0.404 \ \mathrm{d}$, respectively, while $C$, $C_\mathrm{A}$, $\phi_\mathrm{A}$, $C_\mathrm{B}$, and $\phi_\mathrm{B}$ are the free parameters for fitting. 
The original light curve and the fitted long-term trend are illustrated in Fig. \ref{TESS_lc_method}a. 
After fitting the long-term trend, we calculate the normalized flux contributed by the flare by 
\begin{equation}
\Delta F_\mathrm{flare}(t) = \frac{F_\mathrm{TESS}(t) - F_\mathrm{TESS,trend}(t)}{C}, 
\end{equation}
where $C$ is the fitting result of the parameter $C$ in Equation (\ref{periodic_function}) and stands for the average quiescent flux of this binary system. 
The normalized light curve in Fig. \ref{TESS_lc+FAST_ds}a, b, and Fig. \ref{TESS_lc_method}b stands for $\left( \Delta F_\mathrm{flare}(t) + 1 \right)$. 
$\Delta F_\mathrm{flare} (t)$ is then processed by a Gaussian filter (time as the abscissa) with a $\text{3-sigma}$ (i.e., $3 \times 20 \ \mathrm{s}$) Gaussian kernel, resulting in a smoothed light curve $\Delta F'_\mathrm{flare} (t)$ solely to identify the start time and end time of this superflare. 
This smoothed light curve is illustrated in Fig. \ref{TESS_lc_method}b. 
The start time $t_\mathrm{start}$ and end time $t_\mathrm{end}$ are then defined as the first and last time points with $\Delta F'_\mathrm{flare} (t)$ greater than $3$ times the relative average photometric error ($3 \times 6.5 \times 10^{-4}$) provided by the TESS pipeline, which are $3130 \ \mathrm{s}$ and $17130 \ \mathrm{s}$ since $15:23:20 \ \mathrm{UTC}$ on Sep. 8th, 2022, respectively. 
The peak time $t_\mathrm{peak}$ is defined as the time point when $\Delta F'_\mathrm{flare} (t)$ reaches its maximum, which is $3500 \ \mathrm{s}$ since $15:23:20 \ \mathrm{UTC}$ on Sep. 8th, 2022. 

\indent The calculations of the flare area and bolometric energy basically follow the methodology introduced by \citet{Shibayama2013ApJS209.5}. 
Here we briefly summarize the main procedures in the calculations as follows: 
Assuming that the spectra from both the quiescence and the flare can be described by black-body radiation, the projected flare area is calculated as 
\begin{equation}\label{flare_area_Eq}
A_\mathrm{flare}(t) = 
\Delta F_\mathrm{flare}(t) \pi R_\star^2 \frac{
\int_{\lambda_\mathrm{min}}^{\lambda_\mathrm{max}} R_\mathrm{\lambda} B_\mathrm{\lambda}(T_\mathrm{eff}) \ \mathrm{d}\lambda}{
\int_{\lambda_\mathrm{min}}^{\lambda_\mathrm{max}} R_\mathrm{\lambda} B_\mathrm{\lambda}(T_\mathrm{flare}) \ \mathrm{d}\lambda}, 
\end{equation}
where $\lambda_\mathrm{min}$ and $\lambda_\mathrm{max}$ are the lower and upper limits of the TESS bandpass, which are approximately $6000 \ \mathrm{\AA}$ and $10000 \ \mathrm{\AA}$, respectively, 
$R_\lambda$ is the response function of the TESS instrument \citep{Ricker2015JATIS1.014003}, 
$B_\lambda$ is the Planck function, 
$T_\mathrm{eff}$ is the effective temperature of the target star ($3585 \ \mathrm{K}$ for EQ Peg A and $3309 \ \mathrm{K}$ for EQ Peg B \citep{Gaia_Collaboration2018AA616.A1}), 
$T_\mathrm{flare}$ is the effective temperature of the flare.  
We adopt a $T_\mathrm{flare}$ of $10000 \ \mathrm{K}$, which is a commonly accepted estimate for stellar flares \citep[e.g.,][]{Mochnacki1980ApJ239.L27,Hawley1992ApJS78.565,Shibayama2013ApJS209.5,Namekata2017ApJ851.91,Namekata2021NA6.241}. 
To ensure a conservative estimate of the flare's radiated energy, we also consider a relatively large uncertainty of $2000 \ \mathrm{K}$ in $T_\mathrm{flare}$. 
The bolometric luminosity of the flare is then calculated using the Stefan-Boltzmann law as 
\begin{equation}
L_\mathrm{flare}(t) = \sigma_\mathrm{SB} T_\mathrm{flare}^4 A_\mathrm{flare}(t), 
\end{equation}
where $\sigma_\mathrm{SB}$ is the Stefan-Boltzmann constant. 
And the bolometric energy of the flare is calculated by integrating the bolometric luminosity from $t_\mathrm{start}$ to $t_\mathrm{end}$ as 
\begin{equation}
E_\mathrm{flare,bol} = \int_{t_\mathrm{start}}^{t_\mathrm{end}} L_\mathrm{flare}(t) \ \mathrm{d}t. 
\end{equation}

\subsection{FAST data analysis}\label{section_FAST_data_analysis}
\indent The retrieval of the FAST dynamic spectra and the subsequent analyses include the following key procedures: 

\indent \textbf{Flux calibration.} 
Our observation included periodic noise injections from the noise diode in order to perform the flux and polarization calibrations. 
The sampling time in our observation is $196.608 \ \mathrm{\mu s}$. 
The noise temperature $T_\mathrm{noise}(f)$ is known at each observation frequency channel (channel width $0.488 \ \mathrm{MHz}$), and was set to $\sim 10 \ \mathrm{K}$ and injected for $5120$ sampling times ($\sim 1 \ \mathrm{s}$) in every $10240$ sampling times ($\sim 2 \ \mathrm{s}$). 
The time intervals with and without the noise injections are denoted as noise \texttt{ON} and noise \texttt{OFF} intervals, respectively. 
At each frequency channel, the signals expressed in the digital outputs $N$ are averaged over the noise \texttt{ON} and noise \texttt{OFF} intervals, respectively, and the difference between these two averages is considered to be the signals contributed solely by the noise diode. 
Then, we derive the conversion coefficient $k$ from the digital output to the antenna temperature $T_\mathrm{A}$ by comparing the noise signals with the known noise temperatures at each frequency channel, and we obtain 
\begin{equation}
T_\mathrm{A}(f,t) = k(f) N(f,t). \label{T_A}
\end{equation}
This procedure is performed for every single Flexible Image Transport System (FITS) file. 
The calculation of this conversion coefficient is based solely on Stokes $I$ data, but it can be used for the flux calibration of all four Stokes parameters, i.e., Stokes $I, Q, U, V$. 
For noise \texttt{OFF} intervals, the dynamic spectra for further analysis are calculated directly from Equation (\ref{T_A}); 
whereas for noise \texttt{ON} intervals, the noise temperature $T_\mathrm{noise}(f)$ is subtracted after performing Equation (\ref{T_A}) to produce the dynamic spectra for further analysis. 

\indent \textbf{Polarization calibration.} 
In an ideal situation, the noise diode signals should be $100\%$ linearly polarized, which can be expressed with Stokes parameters as $I=U$ and $Q=V=0$. 
However, due to the instrumental response and polarization leakage, the four Stokes parameters can influence each other, which can be characterized by the Mueller matrix \citep[e.g.,][]{Britton2000ApJ532.1240,van-Straten2004ApJS152.129,Sun2021RAA21.282,Zhang2023ApJ953.65} as shown in Equation (\ref{Mueller}), 
\begin{figure*}
\centering
\begin{equation} \label{Mueller}
\begin{pmatrix}
I' \\ Q' \\ U' \\ V' 
\end{pmatrix} = 
\begin{pmatrix}
\cosh{\left( 2 \tilde{\gamma} \right)} & \sinh{\left( 2 \tilde{\gamma} \right)} & 0 & 0 \\
\sinh{\left( 2 \tilde{\gamma} \right)} & \cosh{\left( 2 \tilde{\gamma} \right)} & 0 & 0 \\ 
0 & 0 &  \cos{\left( 2 \tilde{\varphi} \right)} & \sin{\left( 2 \tilde{\varphi} \right)} \\
0 & 0 & -\sin{\left( 2 \tilde{\varphi} \right)} & \cos{\left( 2 \tilde{\varphi} \right)}
\end{pmatrix}
\begin{pmatrix}
I \\ Q \\ U \\ V 
\end{pmatrix}. 
\end{equation}
\end{figure*}
where $\left( I,Q,U,V \right)$ are the real Stokes parameters, $\left( I',Q',U',V' \right)$ are the detected ones, $\tilde{\gamma}$ denotes the differential gain, and $\tilde{\varphi}$ denotes the differential phase. 
The detected Stokes parameters contributed solely by the noise diode are calculated using the same method mentioned in the last step, i.e., they are averaged over the noise \texttt{ON} and noise \texttt{OFF} intervals, respectively, and the differences between noise \texttt{ON} and noise \texttt{OFF} are then used to represent $\left( I',Q',U',V' \right)$ contributed solely by the noise diode. 
For the signals from the noise diode, we assume $I=U$ and $Q=V=0$, then $\tilde{\gamma}$ and $\tilde{\varphi}$ are solved via Equation (\ref{Mueller}). 
The real Stokes parameters $\left( I,Q,U,V \right)$ at each time and frequency point can then be calculated with $\tilde{\gamma}$, $\tilde{\varphi}$, and $\left( I',Q',U',V' \right)$. 

\begin{figure*}
\centering
\includegraphics[width=0.9\textwidth]{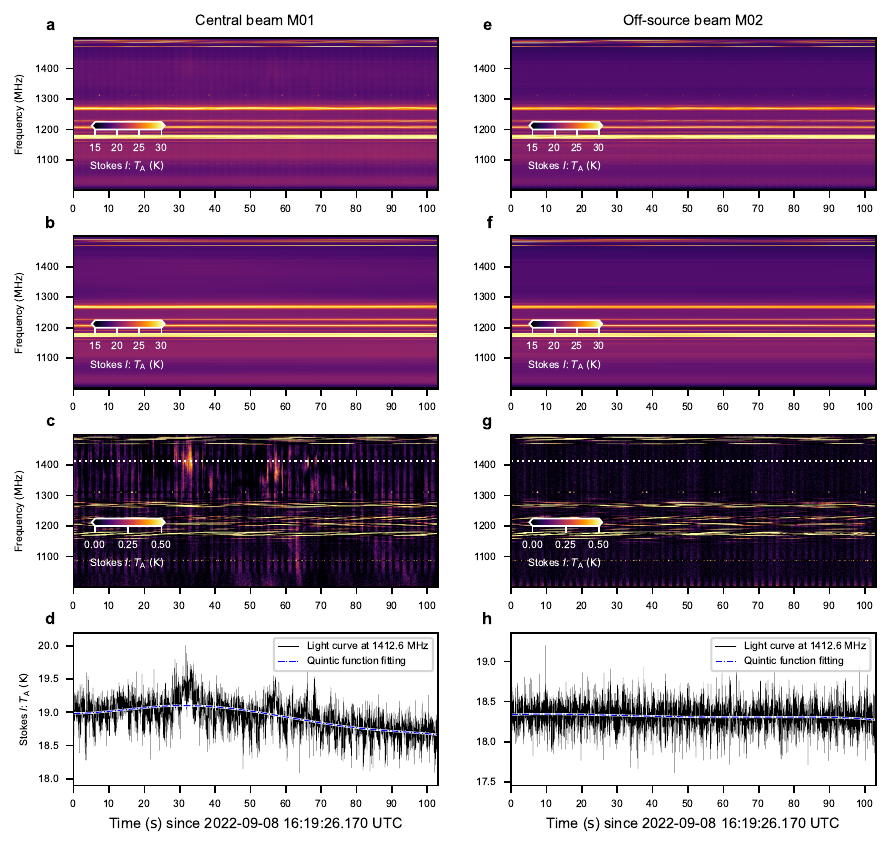}
\caption{Process of slowly-varying background removal within FAST dynamic spectra. 
(a) The FAST dynamic spectrum of Stokes $I$ from the central beam (on-source beam) M01 before background removal. 
(b) The quintic polynomial fitting result of each frequency channel in (a). 
(c) The FAST dynamic spectrum after the quintic polynomial correction, i.e. subtracting data in (b) from data in (a). 
These data are still before the broad-band fast-varying background removal. 
(d) An example of the light curve before background removal and the corresponding quintic polynomial fitting result, which are retrieved from the certain frequency channel whose central frequency is illustrated by the horizontal white dotted line in (c). 
(e)-(h) Similar to (a)-(d), but for an off-source beam M02. 
All panels show the same time range as Fig. \ref{FAST_ds}a. 
}
\label{ds_background_removal_process}
\end{figure*}

\indent \textbf{Background removal.} 
For every single FAST FITS file, the signals of both Stokes $I$ and $V$ at each frequency channel are fitted with a quintic function. 
Then, the fitting results are subtracted from both the Stokes $I$ and $V$ data to remove the slowly-varying background of the two Stokes parameters, and this correction is done independently per frequency channel, as shown in Fig. \ref{ds_background_removal_process}. 
Additionally, the $40 \ \mathrm{MHz}$ wide frequency band with the lowest flux level and without significant RFIs is further subtracted from the Stokes $I$ data to remove the special broad-band fast-varying background of Stokes $I$. 
Such a broad-band fast-varying background of Stokes $I$, which can be seen more clearly in Fig. \ref{ds_background_removal_process}g from an off-source beam M02 (see \citet{Jiang2020RAA20.064J} for the numbering of the FAST 19 beams), is likely contributed by the fluctuating system temperature \citep{Zhang2023ApJ953.65}. 
In the ideal case, after the two-step background removal procedure described above, the distribution of Stokes $I$ of each pixel within the quiescent region of the dynamic spectrum would be expected to peak near $0 \ \mathrm{K}$. 
In practice, however, the peaks of the corresponding distributions are found to be $\sim -0.02 \ \mathrm{K}$ and $\sim -0.06 \ \mathrm{K}$ after each background removal step, respectively. 
Consequently, for signals exceeding $5\sigma$ (i.e. the radio spikes identified in this work), the fractions of real signal that could have been removed by the two steps are estimated to be $\lesssim 9\%$ and $\lesssim 18\%$, respectively. 
Pixels with negative Stokes $I$ values are excluded from further analysis. 

\indent \textbf{RFI flagging.} 
The radio bursts are mainly observed within $1300 - 1500 \ \mathrm{MHz}$, among which we choose an uncontaminated frequency range of $1317 - 1464 \ \mathrm{MHz}$ for further analysis, since neither known nor observable RFIs exist within this frequency range. 

\begin{figure}
\centering
\includegraphics[width=0.45\textwidth]{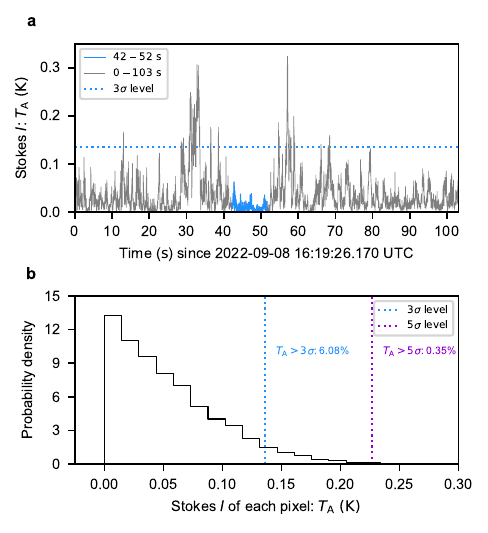}
\caption{(a) The light curve of FAST Stokes $I$, which is retrieved by averaging the dynamic spectrum shown in Fig. \ref{FAST_ds}a over frequency channels corresponding to the quiescent region, i.e. $1322 - 1454 \ \mathrm{MHz}$. 
The time interval corresponding to the quiescent region, i.e. $42 - 52 \ \mathrm{s}$, is highlighted in blue on the light curve. 
(b) Histogram of Stokes $I$ of each pixel within the quiescent region of the dynamic spectrum. 
The blue and violet vertical dotted lines illustrate the $3\sigma$ and $5\sigma$ levels, respectively. 
The fractions of pixels with $T_\mathrm{A} > 3\sigma$ and $T_\mathrm{A} > 5\sigma$ in the quiescent region are $6.08\%$ and $0.35\%$, respectively. 
}
\label{background_lc+hist}
\end{figure}

\begin{figure}
\centering
\includegraphics[width=0.5\textwidth]{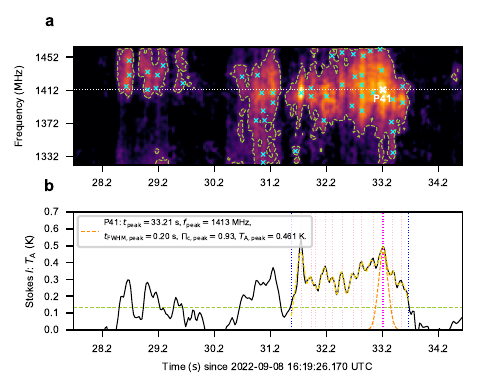}
\caption{Example of the analysis of the radio spikes in the FAST dynamic spectra. 
(a) A zoom-in of the FAST Stokes $I$ dynamic spectrum with a Gaussian filter. 
The magnified white cross denote the 41st spike (denoted as $\text{P41}$) which is used for analysis in this example. 
The meanings of other elements in this panel are the same as those in Fig. \ref{FAST_ds}f. 
(b) The light curve of Stokes $I$ retrieved from the frequency channel of $\text{P41}$. 
The two blue vertical dotted lines suggest the boundaries of the neighborhood of $\text{P41}$ above $3\sigma$ level. 
The yellow dashed line illustrates the multi-Gaussian fitting result of the neighborhood of $\text{P41}$, and the pink vertical dotted line suggests the spikes identified within this neighborhood, which are used as the initial values for the mean of each Gaussian component in the fitting. 
The orange dashed line illustrates the single Gaussian component representing $\text{P41}$, and the magenta vertical dotted line suggests the location of $\text{P41}$ in the time domain, which is used as the initial value for this certain Gaussian component in the fitting. 
The key parameters of $\text{P41}$ obtained from the fitting result are indicated in the panel legend. 
The meanings of other elements in this panel are the same as those in Fig. \ref{FAST_ds}j. 
}
\label{FAST_ds_fs}
\end{figure}

\indent \textbf{Analysis of fine structures.} 
The FAST dynamic spectra are rebinned by a factor of $128$ in time and $4$ in frequency, which yields a time resolution of $25.16 \ \mathrm{ms}$ and a frequency resolution of $1.95 \ \mathrm{MHz}$ to increase the SNR. 
Since different beams exhibit different quiescent noise levels even within the same observing period \citep{Jiang2020RAA20.064J}, we select a quiescent segment adjacent to the radio bursts in the dynamic spectrum of the on-source beam M01 to estimate the corresponding noise level. 
A quiescent part of the dynamic spectrum (
$42 - 52 \ \mathrm{s}$, $1322 - 1454 \ \mathrm{MHz}$), 
as shown in Fig. \ref{FAST_ds}a, 
gives the standard deviation in the quiescence $\sigma=0.045 \ \mathrm{K}$ in antenna temperature. 
The Stokes $I$ value of each pixel within the quiescent region exhibits a monotonically decreasing distribution toward larger values, as shown in Fig. \ref{background_lc+hist}b. 
Since the fraction of pixels with $T_\mathrm{A} > 5\sigma$ in the quiescent region is only $0.35\%$, the $5\sigma$ level can be regarded as a reasonable threshold for identifying statistically significant signals. 
We note that the noise distribution of the quiescent region is highly non-gaussian (see Fig. \ref{background_lc+hist}b), and therefore our true threshold corresponding to false-alarm rate $<0.35\%$ is approximately $3\sigma$, 
which remains a reasonable detection threshold in the presence of this non-gaussian noise. 
In order to distinguish the fine structures within the dynamic spectra more effectively, the dynamic spectra are further processed by a Gaussian filter (time and frequency as the abscissae) with a $\text{1-sigma}$ (i.e., $25.16 \ \mathrm{ms}$ and $1.95 \ \mathrm{MHz}$, respectively) Gaussian kernel, resulting in smoothed dynamic spectra. 
A $7 \ \text{pixels} \times 7 \ \text{pixels}$ frame is scanned over the entire dynamic spectrum. 
The central pixel is marked as a local peak $(t_\mathrm{peak},f_\mathrm{peak})$ if it has the highest flux level within the frame and if the flux is also greater than $5 \sigma$. 
Totally $87$ such local peaks (denoted as $\text{P1} - \text{P87}$) are identified within the FAST dynamic spectrum of Stokes $I$. 
Multi-Gaussian fitting is performed on the light curve retrieved at the frequency channel of each local peak, as shown in Fig. \ref{FAST_ds_fs}b. 
Only the neighborhood above the $3\sigma$ level is taken into consideration, and the number of Gaussian functions depends on the number of peaks within this neighborhood. 
The FWHM of the single Gaussian component representing the certain local peak is regarded as the duration of each spike (denoted as $t_\mathrm{FWHM,peak}$), 
while the peak value of this Gaussian component is regarded as the antenna temperature of the certain local peak (denoted as $T_\mathrm{A,peak}$). 
The degree of circular polarization of each local peak $\mathit{\Pi}_\mathrm{c,peak}$ is calculated by dividing the Stokes $V$ value by the Stokes $I$ value at $(t_\mathrm{peak},f_\mathrm{peak})$. 
Note that Stokes $V$ shows discontinuity in the frequency domain above $1453 \ \mathrm{MHz}$, as shown in Fig. \ref{TESS_lc+FAST_ds}d. 
These data are unreliable and, therefore, are excluded when calculating the degrees of circular polarization of the local peaks. 

\indent \textbf{Estimation of brightness temperatures.} 
The projected flare area $A_\mathrm{flare}(t)$ at each time during the white-light flare has been calculated in the TESS data analysis.  
Since, on the one hand, not all flare loops within the flaring region are capable of producing ECM emission, and on the other hand, even within the ECM-producing loops, the loss-cone- or horseshoe-distributed electrons are expected to exist only at specific locations \citep[e.g.,][]{Treumann2006AAR13.229}, 
we take $A_\mathrm{flare}$ at the peak of the white-light flare as an upper limit of the projected radio source area $A_\mathrm{RB}$. 
Considering the Rayleigh-Jeans law, the observed flux densities of the local peaks can be expressed as
\begin{equation}
F_\mathrm{peak} = \frac{2 k_\mathrm{B} f_\mathrm{peak}^2 T_\mathrm{B,peak}}{c^2} \omega_\mathrm{RB}, 
\end{equation}
where $T_\mathrm{B,peak}$ is the brightness temperatures of the local peaks, $\omega_\mathrm{RB}$ is the solid angle of the radio source relative to the observer, $k_\mathrm{B}$ is the Boltzmann constant, and $c$ is the speed of light. 
Since 
\begin{equation}
\omega_\mathrm{RB} = \frac{A_\mathrm{RB}}{d^2}, 
\end{equation}
where $d = 6.26 \ \mathrm{pc}$ is the distance of EQ Peg \citep{Gaia_Collaboration2023AA674.A1}, 
we have
\begin{equation}
\begin{aligned}
F_\mathrm{peak} &= 
\frac{2 k_\mathrm{B} f_\mathrm{peak}^2 T_\mathrm{B,peak}}{c^2} \frac{A_\mathrm{RB}}{d^2} \\
&\lesssim 
\frac{2 k_\mathrm{B} f_\mathrm{peak}^2 T_\mathrm{B,peak}}{c^2} \frac{A_\mathrm{flare}}{d^2}. 
\end{aligned}
\end{equation}
Finally, the brightness temperatures of the local peaks $T_\mathrm{B,peak}$ are estimated by 
\begin{equation}\label{brightness_temperature_Eq}
T_\mathrm{B,peak} \gtrsim \frac{c^2 d^2}{2 k_\mathrm{B} f_\mathrm{peak}^2 A_\mathrm{flare}} F_\mathrm{peak}. 
\end{equation}

\indent Moreover, in order to testify the reliability of taking $A_\mathrm{flare}$ as an upper limit of $A_\mathrm{RB}$, we also attempt to directly estimate the radio source area based on radio observations. 
\citet{Benz1985SP96.357} suggests that the source dimension $l_\mathrm{RB}$ of a radio spike is determined by its bandwidth as
\begin{equation}
l_\mathrm{RB} \sim \lambda_{B} \frac{\Delta f}{f_\mathrm{peak}}, 
\end{equation}
where $\lambda_{B}$ stands for the magnetic scale length, which is approximately $\lambda_{B} \sim r/3$ under the dipole situation, and $r$ is the distance between the radio source and the center of the star. 
Considering that $f_\mathrm{peak} \sim 1398 \ \mathrm{MHz}$ corresponds to $r \sim 2 R_\star$ for both components under the dipole field assumption, and $\Delta f / f_\mathrm{peak} \lesssim 0.04$ as mentioned in Observations, we have 
\begin{equation}
\frac{\pi\left(l_\mathrm{RB}/2\right)^2}{\pi R_\star^2} \lesssim 2 \times 10^{-4} 
\end{equation}
for both EQ Peg A and B, which are located within the range estimated from the projected flare area, i.e., $A_\mathrm{RB,A} / \pi R_\mathrm{\star,A}^2 \lesssim 1.2 \times 10^{-3}$ for EQ Peg A or 
$A_\mathrm{RB,B} / \pi R_\mathrm{\star,B}^2 \lesssim 8.1 \times 10^{-4}$ for EQ Peg B. 
Recent solar imaging-spectroscopy observations \citep[e.g.,][]{Ma2026NC17.5131} indicate that spike-like radio bursts originate from compact source regions associated with localized energy release, suggesting a characteristic source scale comparable to that of the underlying flare energy-release sites. 
Such consistencies reinforce the validity of the former estimation strategy.


\begin{figure*}
\centering
\includegraphics[width=1.0\textwidth]{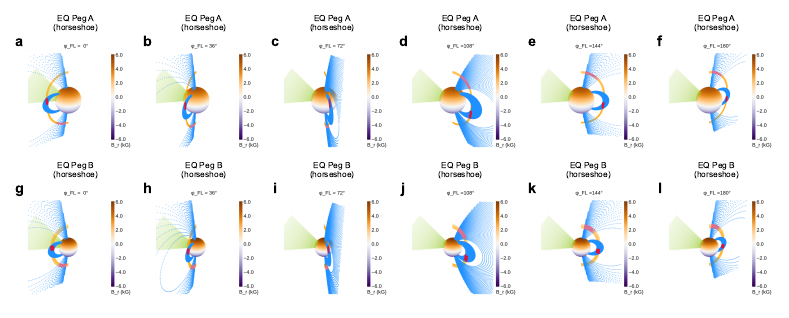}
\caption{Snapshots of the emission sources for different azimuthal angles of the field lines ($\varphi_\mathrm{FL}$), under the condition of horseshoe-distributed electrons. 
The meanings of all the elements in this figure are the same as those in Fig. \ref{dipole_3d}a and c. 
(a)-(f) Emission sources as for EQ Peg A when $\varphi_\mathrm{FL} = 0^\circ$, $36^\circ$, $72^\circ$, $108^\circ$, $144^\circ$, and $180^\circ$, respectively. 
(g)-(l) Similar to (a)-(f), but for the situations of EQ Peg B. 
}
\label{dipole_3d_snapshots_appendix_horseshoe}
\end{figure*}

\begin{figure*}
\centering
\includegraphics[width=1.0\textwidth]{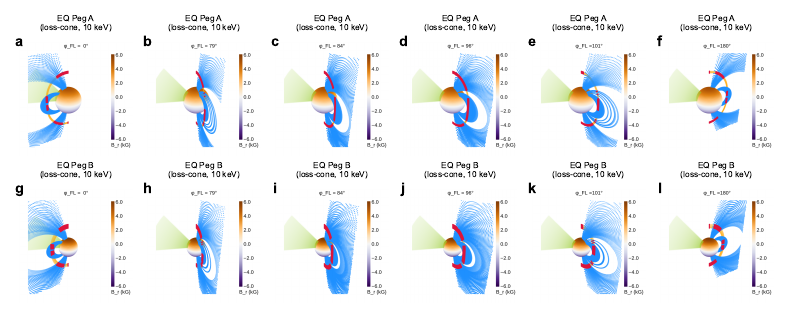}
\caption{Similar to Fig. \ref{dipole_3d_snapshots_appendix_horseshoe}, but under the condition of loss-cone-distributed electrons with $10 \ \mathrm{keV}$ kinetic energy, 
and for $\varphi_\mathrm{FL} = 0^\circ$, $79^\circ$, $84^\circ$, $96^\circ$, $101^\circ$, and $180^\circ$, respectively. 
The meanings of all the elements in this figure are the same as those in Fig. \ref{dipole_3d}b and d. 
}
\label{dipole_3d_snapshots_appendix_loss-cone}
\end{figure*}

\begin{figure}
\centering
\includegraphics[width=0.45\textwidth]{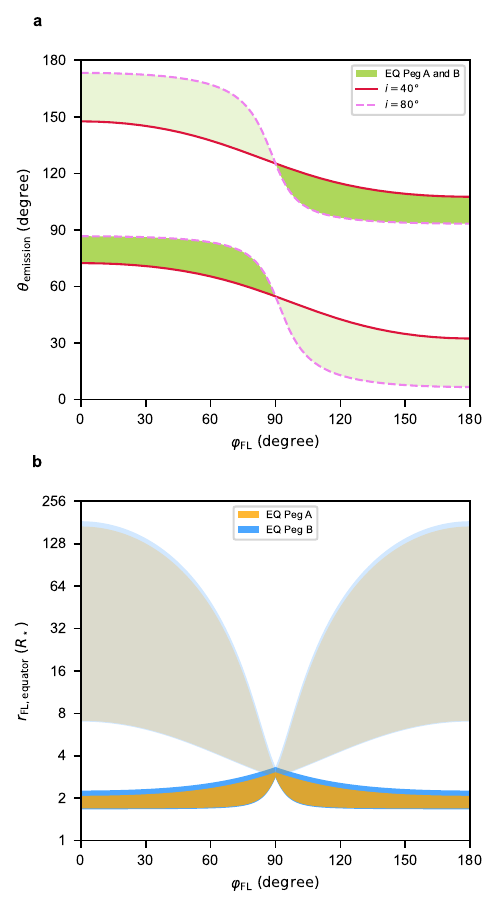}
\caption{Distributions of the emission sources driven by horseshoe-distributed electrons. 
(a) Polar angles of the emission sources ($\theta_\mathrm{emission}$) for field lines with different azimuthal angles. 
The yellow-green shaded area indicates the $\theta_\mathrm{emission}$ values corresponding to $i$ between $40^\circ$ and $80^\circ$, while the red solid line and the violet dashed line indicate the $\theta_\mathrm{emission}$ values for the special situations of $i = 40^\circ$ and $i = 80^\circ$, respectively. 
The pattern shown in this panel holds for both components, i.e. EQ Peg A and B. 
(b) Distances between the top of the field lines passing through the emission sources and the center of the star ($r_\mathrm{FL,equator}$) for field lines with different azimuthal angles. 
The orange shaded area indicates the $r_\mathrm{FL,equator}$ values 
in the unit of $R_{\star,\mathrm{A}}$ as for EQ Peg A, 
while the blue shaded area indicates the situation 
in the unit of $R_{\star,\mathrm{B}}$ as for EQ Peg B. 
The darker yellow-green, orange, and blue elements in the panels refer to the more likely sources than the lighter ones (see Section \ref{section_determination} Determination of sources of radio spikes for details). 
Note that, the potential horseshoe-driven radio sources shrink into one point when $\varphi_\mathrm{FL} = 90^\circ$, which is nearly an impossible solution from a physical perspective. 
This indicates that it is unlikely for the radio source to be located on the magnetic field lines with $\varphi_\mathrm{FL} = 90^\circ$. 
Also see animated Fig. \ref{video_s1} online. 
}
\label{dipole_3d_statistics_appendix}
\end{figure}

\begin{figure}
\centering
\includegraphics[width=0.45\textwidth]{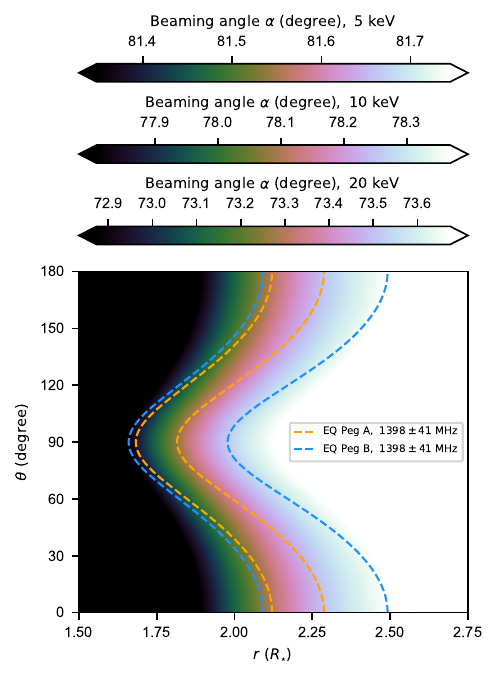}
\caption{Spatial distributions of the beaming angle for ECM driven by loss-cone-distributed electrons, i.e. $\alpha(r,\theta)$. 
The horizontal and vertical axes stand for the distance to the center of the star ($r$) and the polar angle ($\theta$), respectively (see Section \ref{section_determination} Determination of sources of radio spikes for the detailed definition of this spherical coordinate system). 
The three colorbars, from top to bottom, denote the beaming angle $\alpha$ for ECM driven by loss-cone-distributed electrons with kinetic energies of $5,\ 10,\ 20 \ \mathrm{keV}$, respectively. 
The orange and blue dashed lines indicate the boundaries of regions corresponding to $1398 \pm 41 \ \mathrm{MHz}$ for EQ Peg A and B, respectively. 
}
\label{loss-cone_beaming_angle}
\end{figure}

\begin{figure*}
\begin{interactive}{animation}{Video_S1.mp4}
\includegraphics[width=0.8\textwidth]{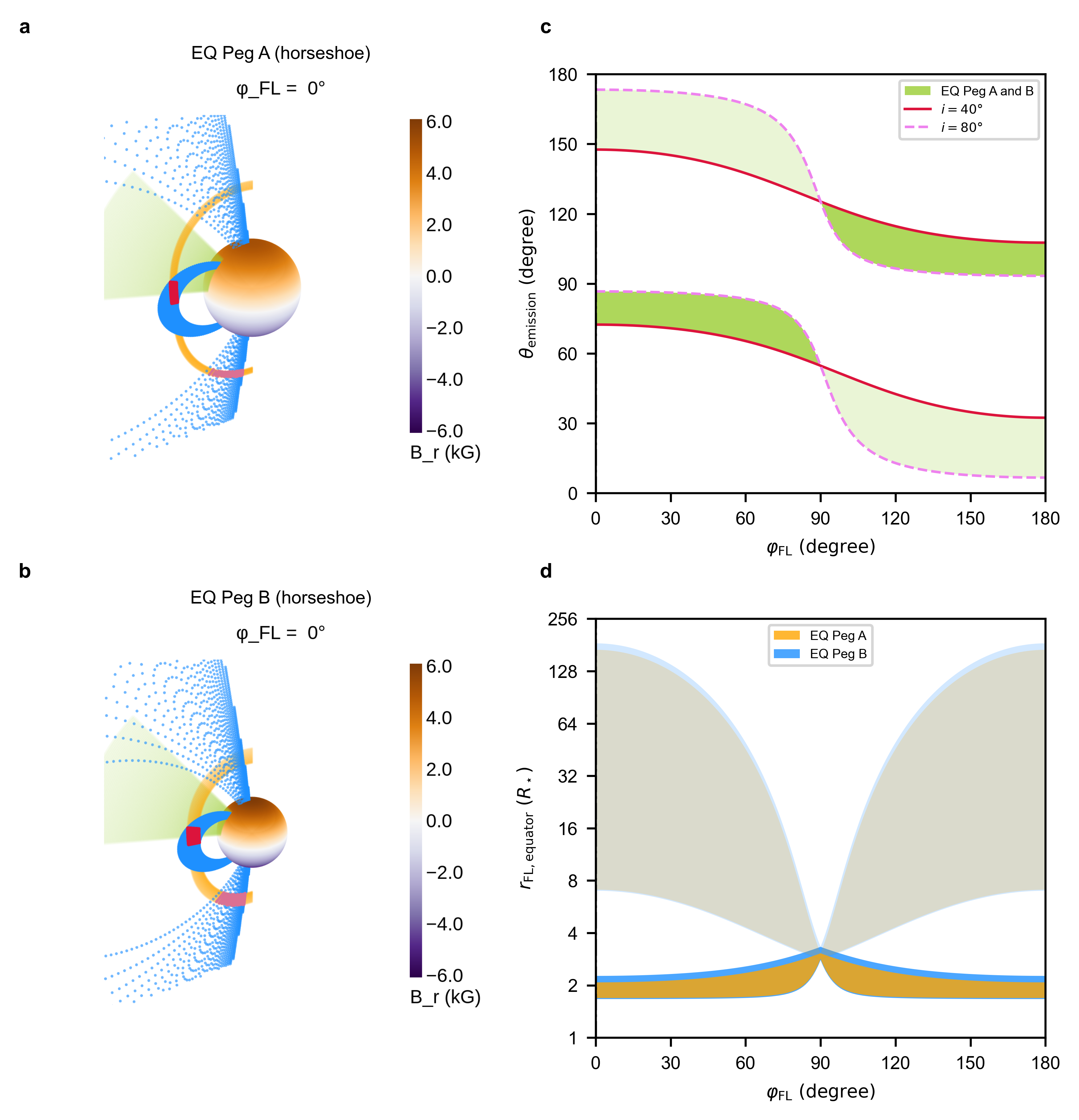}
\end{interactive}
\caption{This figure is available as an animation online, which lasts for about $18 \ \mathrm{s}$, and illustrates the possible radio sources driven by horseshoe-distributed electrons, including all cases for azimuthal angles $\varphi_\mathrm{FL}$ ranging from $0^\circ$ to $180^\circ$. 
The static representation shows the first frame of the animation. 
Please see Section \ref{section_determination} and Fig. \ref{dipole_3d} for a detailed description of this animation. }
\label{video_s1}
\end{figure*}

\begin{figure}
\begin{interactive}{animation}{Video_S2.mp4}
\includegraphics[width=0.4\textwidth]{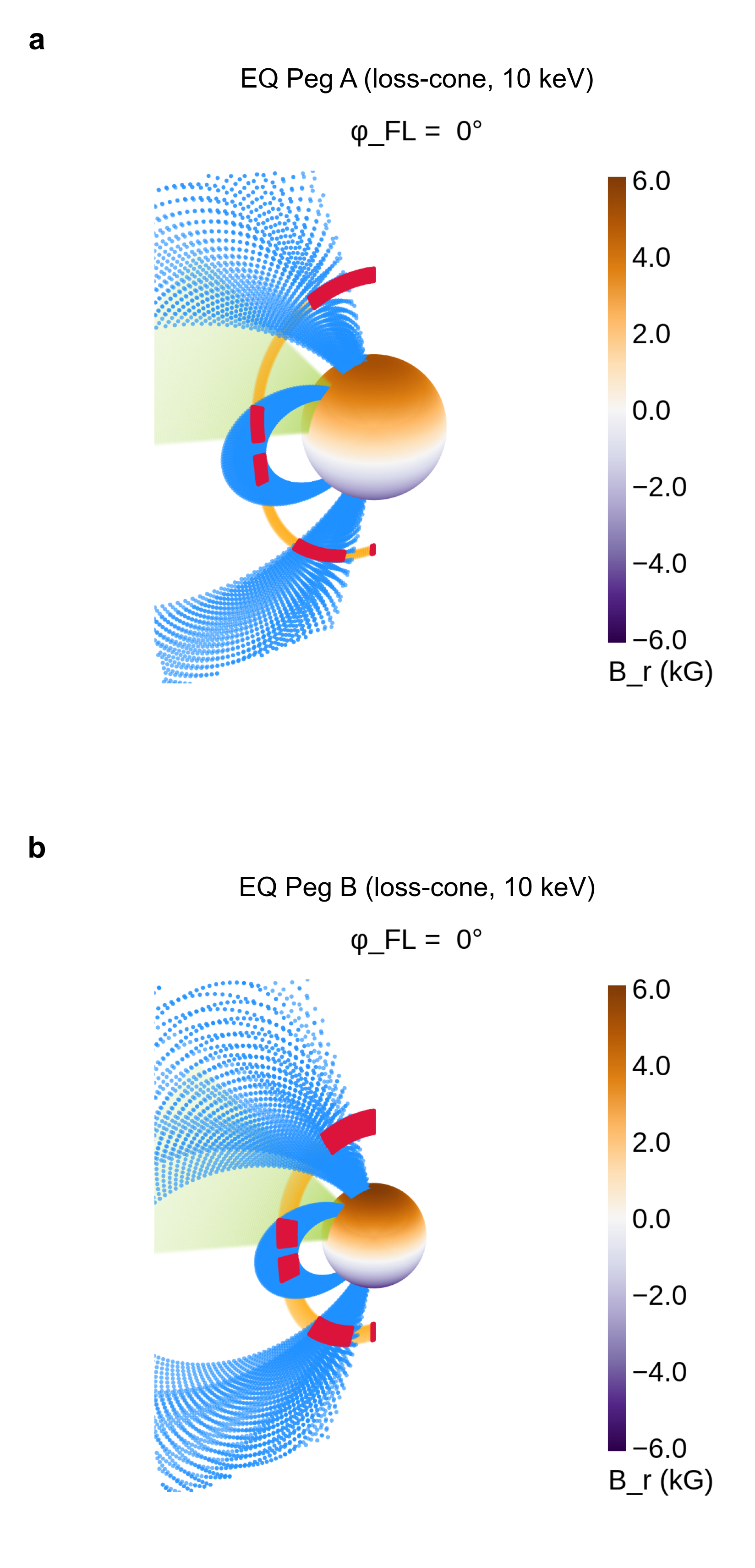}
\end{interactive}
\caption{This figure is available as an animation online, which lasts for about $18 \ \mathrm{s}$, and illustrates the possible radio sources driven by loss-cone-distributed electrons, including all cases for azimuthal angles $\varphi_\mathrm{FL}$ ranging from $0^\circ$ to $180^\circ$. 
The static representation shows the first frame of the animation. 
Please see Section \ref{section_determination} and Fig. \ref{dipole_3d} for a detailed description of this animation. }
\label{video_s2}
\end{figure}

\bibliography{sample701}{}
\bibliographystyle{aasjournalv7}

\end{document}